\documentclass[aps,prc,twocolumn,superscriptaddress,nofootinbib]{revtex4-1}

\usepackage{amssymb}
\usepackage{amsmath}
\usepackage{siunitx}
\usepackage{pifont}
\usepackage{multirow}
\usepackage{color}

\usepackage{hyperref}
\hypersetup{letterpaper, colorlinks=true, breaklinks=True, citecolor=blue}

\usepackage{nicefrac}

\usepackage{graphicx}

\makeatletter

\newcommand{\Rmnum}[1]{\expandafter\@slowromancap\romannumeral #1@}
\makeatother

\newcommand{\nuc}[2]{${}^{#1}\mathrm{#2}$}

\begin{document}




\title{Study of (${}^{6}$Li, $d$) and (${}^{6}$Li, $t$) reactions on ${}^{22}$Ne and implications for $s$-process nucleosynthesis}


\author{S. Ota}
\email[]{shuyaota@comp.tamu.edu}
\affiliation{Cyclotron Institute, Texas A\&M University, College Station, TX 77843, USA}
\affiliation{NuGrid Collaboration, \url{http://nugridstars.org}}
\author{G. Christian}
\affiliation{Cyclotron Institute, Texas A\&M University, College Station, TX 77843, USA}
\affiliation{Department of Astronomy \& Physics, Saint Mary's University, Halifax, NS B3H~3C3, Canada}
\affiliation{Department of Physics \& Astronomy, Texas A\&M University, College Station, TX 77843, USA}
\affiliation{Nuclear Solutions Institute, Texas A\&M University, College Station, TX 77843, USA}
\author{W. N. Catford}
\affiliation{Department of Physics, University of Surrey, Guildford GU2 7XH, UK}
\author{G. Lotay}
\affiliation{Department of Physics, University of Surrey, Guildford GU2 7XH, UK}
\author{M. Pignatari}
\affiliation{Joint Institute for Nuclear Astrophysics - Center for the Evolution of the Elements, East Lansing, 48823, USA}
\affiliation{E. A. Milne Centre for Astrophysics, Department of Physics and Mathematics, University of Hull, Hull HU6 7RX, UK}
\affiliation{Konkoly Observatory, Research Centre for Astronomy and Earth Sciences, Hungarian Academy of Sciences, Konkoly Thege M. t 15-17, 1121, Budapest, Hungary}
\affiliation{NuGrid Collaboration, \url{http://nugridstars.org}}
\author{U. Battino}
\affiliation{School of Physics and Astronomy, University of Edinburgh, EH9 3FD, UK}
\affiliation{NuGrid Collaboration, \url{http://nugridstars.org}}
\author{E. A. Bennett}
\affiliation{Cyclotron Institute, Texas A\&M University, College Station, TX 77843, USA}
\affiliation{Department of Physics \& Astronomy, Texas A\&M University, College Station, TX 77843, USA}
\author{S. Dede}
\affiliation{Cyclotron Institute, Texas A\&M University, College Station, TX 77843, USA}
\affiliation{Department of Physics \& Astronomy, Texas A\&M University, College Station, TX 77843, USA}
\author{D. T. Doherty}
\affiliation{Department of Physics, University of Surrey, Guildford GU2 7XH, UK}
\author{S. Hallam}
\affiliation{Department of Physics, University of Surrey, Guildford GU2 7XH, UK}
\author{F. Herwig}
\affiliation{Department of Physics and Astronomy, University of Victoria, Victoria, BC V8P5C2, Canada}
\affiliation{Joint Institute for Nuclear Astrophysics - Center for the Evolution of the Elements, East Lansing, 48823, USA}
\affiliation{NuGrid Collaboration, \url{http://nugridstars.org}}
\author{J. Hooker}
\affiliation{Cyclotron Institute, Texas A\&M University, College Station, TX 77843, USA}
\affiliation{Department of Physics \& Astronomy, Texas A\&M University, College Station, TX 77843, USA}
\author{C. Hunt}
\affiliation{Cyclotron Institute, Texas A\&M University, College Station, TX 77843, USA}
\affiliation{Department of Physics \& Astronomy, Texas A\&M University, College Station, TX 77843, USA}
\author{H. Jayatissa}
\affiliation{Cyclotron Institute, Texas A\&M University, College Station, TX 77843, USA}
\affiliation{Department of Physics \& Astronomy, Texas A\&M University, College Station, TX 77843, USA}
\author{A. Matta}
\affiliation{Department of Physics, University of Surrey, Guildford GU2 7XH, UK}
\author{M. Mouhkaddam}
\affiliation{Department of Physics, University of Surrey, Guildford GU2 7XH, UK}
\author{E. Rao}
\affiliation{Cyclotron Institute, Texas A\&M University, College Station, TX 77843, USA}
\affiliation{Department of Physics \& Astronomy, Rutgers University, New Brunswick, NJ, USA}
\author{G. V. Rogachev}
\affiliation{Cyclotron Institute, Texas A\&M University, College Station, TX 77843, USA}
\affiliation{Department of Physics \& Astronomy, Texas A\&M University, College Station, TX 77843, USA}
\affiliation{Nuclear Solutions Institute, Texas A\&M University, College Station, TX 77843, USA}
\author{A. Saastamoinen}
\affiliation{Cyclotron Institute, Texas A\&M University, College Station, TX 77843, USA}
\author{D. Scriven}
\affiliation{Cyclotron Institute, Texas A\&M University, College Station, TX 77843, USA}
\affiliation{Department of Physics \& Astronomy, Texas A\&M University, College Station, TX 77843, USA}
\author{J. A. Tostevin}
\affiliation{Department of Physics, University of Surrey, Guildford GU2 7XH, UK}
\author{S. Upadhyayula}
\affiliation{Cyclotron Institute, Texas A\&M University, College Station, TX 77843, USA}
\affiliation{Department of Physics \& Astronomy, Texas A\&M University, College Station, TX 77843, USA}
\author{R. Wilkinson}
\affiliation{Department of Physics, University of Surrey, Guildford GU2 7XH, UK}



\date{\today}
 
\begin{abstract}

We studied $\alpha$ cluster states in $^{26}$Mg via the $^{22}$Ne($^{6}$Li,$d\gamma$)$^{26}$Mg reaction in inverse kinematics at an energy of $7$ MeV/nucleon. 
States between $E_x$ = 4 - 12 MeV in $^{26}$Mg were populated and relative $\alpha$ spectroscopic factors were determined. 
Some of these states correspond to resonances in the Gamow window of the $^{22}$Ne($\alpha$,n)$^{25}$Mg reaction, 
which is one of the main neutron sources in the astrophysical $s$-process. We show that
$\alpha$-cluster strength of the states analyzed in this work have critical impact on s-process 
abundances. 
Using our new $^{22}$Ne($\alpha$,n)$^{25}$Mg and $^{22}$Ne($\alpha$,$\gamma$)$^{26}$Mg reaction rates, we performed new s-process calculations for massive stars and Asymptotic Giant Branch stars and compared the resulting yields with the yields obtained using other $^{22}$Ne+$\alpha$ rates  from the literature. We observe an impact on the s-process abundances up to a factor of three for intermediate-mass AGB stars and up to a factor of ten for massive stars. 
Additionally, 
states in $^{25}$Mg at $E_x$ $<$ 5 MeV are identified via the $^{22}$Ne($^{6}$Li,$t$)$^{25}$Mg reaction for the first time. 
We present the ($^6$Li, $t$)  
spectroscopic factors of these states and note similarities to the $(d,p$) reaction in terms of reaction selectivity.  

\end{abstract}


\maketitle


\section{Introduction}


Most of the elements heavier than iron are made by
neutron capture processes in stars. 
In the solar system, about half of the abundances of heavy elements are made by the slow neutron-capture process ($s$-process) \cite{Kappeler2011}, while most of the remaining abundances are made by the rapid neutron-capture process ($r$-process) \cite{cowan:21}. Additional contributions from the intermediate neutron-capture processes ($i$-process) \cite{cowan:77} are still a matter of debate \cite{Denissenkov2017}, and explosive nucleosynthesis components from supernovae are expected to be potentially relevant only up to the the Sr-Pd region \cite{roberts:10,arcones:11}.
%
The $r$-process occurs in 
extreme stellar environments such as neutron star mergers \cite{LIGO2017, Drout2017} and rare types of supernovae \cite{cowan:21}. 
Its constituent reactions involve extremely neutron-rich nuclei far from stability \cite{Arnould2007, Sneden2008}.  In contrast, the $s$-process occurs 
during hydrostatic stellar evolution, and because of the lower neutron densities the $s$-process nucleosynthesis path proceeds along the valley of stability.
As a result, the $s$-process 
can potentially be better constrained through accessing the relevant 
nuclear physics, e.g., Ref,~\cite{Bao2000}, 
and stellar computational modeling \cite{Karakas2012}. 
By studying the isotopic pattern of the solar abundances for heavy $s$-process elements, three different $s$-process components have been identified. The main and the strong components are mostly formed in low-to-intermediate mass Asymptotic Giant Branch (AGB) stars in the He-rich intershell region \cite{Straniero2006} ($M/M_\odot \sim 1.5$--$3 $), producing most of the solar $s$-process abundances 
in the $A \sim 90$--$209$ range---i.e., beyond the $N$=50 peak at $^{88}$Sr \cite{gallino:98,kaeppeler:11}.  
The `weak' $s$-process components are  made 
in 
massive stars ($M/M_\odot \gtrsim 8$), during the convective He core 
and  C shell 
burning phases. 
The neutron exposures optained in these conditions produce nuclides in the $A \sim 60$--$90$ mass region, possibly contributing also to the long-lived radioactive isotope $^{60}$Fe 
\cite{rauscher:02,the:07,pignatari:10,pignatari:16}.


The accurate knowledge of 
the rates of neutron-generating reactions 
and of relevant neutron-capture cross sections 
are crucial to move toward 
a complete understanding of $s$-process nucleosynthesis. Along with the ${}^{13}\mathrm{C}(\alpha, n){}^{16}\mathrm{O}$ reaction, $^{22}\mathrm{Ne}(\alpha,n){}^{25}\mathrm{Mg}$ has long been recognized as
one of the 
most important $s$-process neutron sources \cite{peters:68,Iben1975}. For the main $s$-process in AGB stars, most of the neutrons are made by the $^{13}$C($\alpha,n$)$^{16}$O reaction. However, during the Thermal Pulse the $^{22}$Ne($\alpha,n$)$^{25}$Mg reaction is partially activated for temperatures larger than about 0.25 GK, providing an additional neutron exposure and  higher neutron densities \cite{karakas:14}. 
In the He intershell region 
just below the H shell, $^{12}$C is produced via triple-$\alpha$ reactions during the convective Thermal Pulses. 
The $^{12}$C captures protons brought in the He-rich material by the Third-Dredge Up events,  
ultimately forming $^{13}$C following decay of the generated $^{13}$N, and creating the $^{13}$C-pocket, where 
the ${}^{13}\mathrm{C}(\alpha, n){}^{16}$O reaction activates the s-process in radiative conditions \cite{Straniero2006}. 
The following Thermal Pulse mixes the s-process rich material from the ashes of the $^{13}$C-pocket in the He-intershell. The high neutron density generated by the $^{22}$Ne($\alpha,n$)${}^{25}$Mg reaction allows activation of several s-process branching points (e.g., $^{95}$Zr), modifying the isotopic pattern and producing isotopes not accessible during the neutron exposure in the $^{13}$C-pocket \cite{bisterzo:15}. 

In the weak $s$-process, $^{22}$Ne($\alpha,n$)$^{25}$Mg is 
the dominant neutron source throughout both the He core and the C shell over a range of stellar temperatures between about 0.25 GK and 1 GK \cite{raiteri:91,kaeppeler:94,Chieffi1998,pignatari:10}.  
In C-shell burning, 
long-lived $^{60}$Fe 
is also made, thus contributing to the total ejected amount of this isotope \cite{jones:19}.   The production of galactic $^{60}$Fe is of significant interest since it is detected as  a
diffusive $\gamma$-ray source in the Galaxy \cite{Wang2007} and has a signature is identified in the Early Solar System \cite{lugaro:18}. 

After many theoretical and experimental efforts over the past few decades, 
there still exist significant uncertainties in the stellar rate of  $^{22}$Ne($\alpha$,$n$)${}^{25}$Mg, as well as the competing reaction  $^{22}$Ne($\alpha$,$\gamma$)${}^{26}$Mg, e.g.\  as discussed in Ref~\cite{Longland2012}. The uncertainties are dominated by ambiguous strength of some resonances in the Gamow window ($T$=0.2-0.3 GK, corresponding to $E_x$=11.1-11.4 MeV in ${}^{26}$Mg). This is due in large part to difficulties in isolating specific states due to the high level density of $^{26}$Mg in this excitation region, as well as the low overall cross sections which challenge direct measurements.
In the last decade, many new experiments targeting the role of $^{26}$Mg resonances in $^{22}$Ne($\alpha$,$n$)${}^{25}$Mg have been published, building upon the knowledge of resonance properties accumulated in the 1980s and 1990s. These have indicated that the ${}^{22}$Ne$(\alpha,n$)${}^{25}$Mg reaction is dominated by the resonance at $E_x = 11.32$~MeV, and possibly a second resonance in the $E_x \sim 11.15$--11.17 MeV region \cite{Wolke1989, Drotleff1991, Jaeger2001}. 
%
Additionally, Koehler \cite{Koehler2002} pointed out that a $2^+$ resonance just above the neutron threshold, at $E_x=11.11$~MeV may dominate the rate at low temperatures. Massimi \emph{et al}.\ later performed a precise measurement of the neutron and $\gamma$-ray partial widths of this state in a scan of $(n,\gamma$) resonances using a Time-of-Flight neutron beam, confirming Koehler's claim \cite{Massimi2012, Massimi2017}. 

In the 11.15--11.17~MeV region, a candidate $1^-$ state at 11.15~MeV was originally considered as a potential contributor to both ${}^{22}\mathrm{Ne} + \alpha$ capture reactions. Upper limits on $\Gamma_\alpha$ were set both from $({}^6\mathrm{Li},d)$ \cite{Giesen1993} and direct $(\alpha,n)$ \cite{Jaeger2001} measurements. However, this state was later shown to be $1^+$ (and hence non-natural parity) by $(\gamma,\gamma^\prime)$ experiments by Longland \emph{et al}.\ \cite{Longland2009,Longland2012}.
 Talwar \emph{et al}.\ later 
 suggested the presence of a natural parity ($1^-$ or $2^+$) resonance at 11.17 MeV, with a large $\gamma$-ray partial width, indicating an enhanced ($\alpha$,$\gamma$) cross section that suppresses the total neutron production by consuming $^{22}$Ne in competition with the weaker ($\alpha$,$n$) channel \cite{Talwar2016}. Adsley \emph{et al}.\ performed high energy-resolution measurements of $^{26}$Mg($\alpha$,$\alpha$'), $^{26}$Mg($p$,$p$'), and $^{26}$Mg($d$,$d$'), and provided important properties of these resonances above such as excitation energy with high precision \cite{Adsley2017, Adsley2018}. Lotay \emph{et al}.\ observed a strong $\gamma$ transition from the 11.17 MeV state \cite{Lotay2019}. This result was consistent with Talwar \emph{et al}.'s claim of a large $\Gamma_\gamma$, but suggestive of a higher spin state ($J=2$--6). Most recently, Jayatissa \emph{et al}.\ failed to observe a state at 11.17 MeV in a measurement of the $^{22}$Ne($^6$Li,$d$)$^{26}$Mg reaction at sub-Coulomb energies \cite{Jayatissa2019}. 
This reaction mechanism is likely to be capable of populating natural-parity states only with $J \leq 2$, suggesting that the state observed at 11.17 MeV by Talwar \emph{et al}. is high spin ($J \geq 3$). 
All together, the latest experimental results indicate that the 11.17 MeV state has a large $\Gamma_\gamma$ and high spin ($J \geq 3$), thus making its contribution to the s-process negligible.
 
 For the remaining 11.32 MeV resonance, the $(\alpha,\gamma)$ resonance strength is well established, with earlier measurements  \cite{Wolke1989, JaegerPhD} recently confirmed by the new study of Hunt \emph{et al}., which reported $\omega\gamma = 46 \pm 11~\mu$eV. The weighted average of published direct-measurement $(\alpha,\gamma)$ strengths for this resonance is $37 \pm 4~\mu$eV.
 The $(\alpha,n)$ strength of this resonance is more uncertain, with the results of direct measurements in poor statistical agreement \cite{Longland2012}. A recent letter published by the present authors determined the $\Gamma_n$/$\Gamma_\gamma$ branching ratio of this resonance via the $^{22}$Ne($^6\mathrm{Li},d$)$^{26}$Mg reaction in inverse kinematics \cite{Ota2019}. This was done by observing the decay of the recoil $^{26}$Mg into either $^{26}\mathrm{Mg} + \gamma$  or $^{25}\mathrm{Mg} + n$.  Normalizing to the $(\alpha,\gamma)$ strength of $37 \pm 4~\mu$eV, this established an  ($\alpha$,$n$) strength of $42 \pm 11~\mu$eV,  which is a factor ${\sim} 3$ smaller than the past direct measurements.
Based on the results of recent experiments, the 11.32 MeV resonance appears to be the main contributor to the stellar rate across the important temperature range for the $s$-process. At the same time, the 11.17 MeV resonance observed by Talwar \emph{et al}.\  is unlikely to contribute to the stellar reaction.
The main outstanding uncertainty concerns the 11.11 MeV resonance identified by Massimi \emph{et al}.\ \cite{Massimi2017}. 
This state has the potential to dominate the $(\alpha,n)$ rate at low temperatures (below ${\sim} 0.2$~GK). The Monte Carlo rate calculations presented in Ref.~\cite{Ota2019}, which sample the possible strengths from a Porter-Thomas distribution, indicate a large uncertainty in the contribution of this resonance to the stellar rate in this temperature range.

In the present paper, we significantly expand our first letter paper \cite{Ota2019}  wherein we reported $\Gamma_n$/$\Gamma_\gamma = 1.14 \pm 0.26$ at $E_x$=11.32 MeV. In particular, we present a complete description of the experimental setup, which combined the TIARA Si detector array with four HPGe detectors and the MDM spectrometer  at Texas A\&M University. Additionally, we present an analysis of the complete range of strong $\alpha$ cluster states in ${}^{26}$Mg populated in the ${}^{22}\mathrm{Ne}({}^{6}\mathrm{Li},d){}^{26}\mathrm{Mg}$ reaction, as well as states in ${}^{25}$Mg observed in the ${}^{22}\mathrm{Ne}({}^{6}\mathrm{Li},t){}^{25}\mathrm{Mg}$ reaction.
Finally, we present the results of a large scale computer simulation, which determines the effect on s-process abundances of replacing the literature values of the $^{22}\mathrm{Ne}(\alpha,n){}^{25}\mathrm{Mg}$ and $^{22}\mathrm{Ne}(\alpha,\gamma){}^{26}\mathrm{Mg}$ rates with the new rates presented in Ref.~\cite{Ota2019}. 
For the simulations, we used a multi-zone post-processing code, MPPNP \cite{Herwig2018}, to study the sensitivity at various stellar masses and initial metallicities. 
The simulations also demonstrate, in a more general way, the sensitivity of the abundance to the ambiguous reaction rates, by varying the  $\alpha$ cluster strength of relevant ${}^{22}\mathrm{Ne} + \alpha$ resonances.

\section{Experimental Setup}

We performed the experiment at the Texas A\&M University Cycloton Institute, using the K150 cyclotron to produce a beam of 154 MeV $^{22}$Ne$^{(7+)}$ ions. We studied the $({}^{6}\mathrm{Li},d)$ reaction using \nuc{22}{Ne} beam in inverse kinematics, impinging the beam onto an isotopically-enriched 
${}^{6}$LiF target mounted on a 
carbon backing. 
We detected both the deuteron ejectiles and unambiguously identified \nuc{25,26}{Mg} recoils in coincidence, along with the $\gamma$~rays resulting from the de-excitation of states populated in ${}^{25,26}$Mg. Deuterons, \nuc{25,26}{Mg} recoils, and $\gamma$-decays were measured with the TIARA Si detector array, the MDM spectrometer, or an array of four closely-packed HPGe detectors, respectively. A detailed description of each of these components is given in the following sub-sections.

\subsection{TIARA}

TIARA consists of two sets of Si detectors called the ``Hyball'' and ``Barrel'', respectively \cite{Labiche2010}. The Hyball consists of six wedge-shaped, double-sided Si detectors with  16 rings and 8 sectors for each. The effective area of each detector covers an azimuthal angle of $\phi = 54.8^\circ$. Together, the six wedges form an annular shape with inner and outer radii of 32.6 mm and 135.0 mm, respectively.
The Hyball was placed at 15 cm upstream from the target position and thus detected ejectiles emitted at 
laboratory polar angles from $\theta_{lab}$ = 148--168$^\circ$ (note that angles from $\theta_{lab}$ = 138 to $\sim$148$^\circ$ were not illuminated in the present experiment because they were shadowed by the Barrel detector). The Hyball consists of a single active layer of Si with a nominal thickness of  400~$\mu$m.

The Barrel consists of eight resistive charge division Si detectors forming an octagonal barrel around the beam axis, surrounding the target and covering polar angles from $\theta_{lab}$ = 40--145$^\circ$. Each Barrel detector is segmented by four strips in the azimuthal ($\phi$) direction. The z-position (along the beam axis) of incident particles is  determined from the ratio of deposited energies in the upstream and downstream electrodes, after correcting for the ballistic deficit. The z-axis position along the barrel directly corresponds to $\theta_{lab}$ in the present geometry. 
The expected z-position resolution is 1 mm (FWHM) \cite{BrownPhD}, leading to the $\theta_{lab}$ resolution better than 2$^\circ$. 
Further details about energy and angle measurements in the Barrel can be found in, e.g.\ Ref.~\cite{BrownPhD}. 
The resistive-strip layer of the Barrel (``Inner Barrel'') has a nominal thickness of 400~$\mu$m and is backed by an unsegmented outer layer (``Outer Barrel'') with 1~mm active thickness. Thus, particles which punch through the Inner Barrel can be identified by the conventional $\Delta E$-$E$ method (see Figure~\ref{fig:Fig3}b).

From  measurements using multi-nuclide ($^{239}$Pu, $^{241}$Am and $^{244}$Cm) $\alpha$ sources, the intrinsic energy resolutions of the Hyball and Barrel were determined to be 40 and 100 keV FWHM, respectively. These correspond to 180--280 keV and 400--600 keV resolutions in the center-of-mass system for the ${}^{22}\mathrm{Ne}({}^{6}\mathrm{Li}, d)$ and ${}^{22}\mathrm{Ne}(d, p)$ reactions. Center-of-mass resolutions in the barrel are dominated by the poor angular resolution $\delta_{\theta_{CM}} \sim$ 1$^{\circ}$, relative to the steep slope of the $E_{lab}$ vs.\ $\theta_{lab}$ kinematic curves in the corresponding angular region.
As the Barrel's energy resolution is insufficient to separate states in the $^{26}$Mg  
excitation energy spectrum, and furthermore because the $^{26}$Mg in coincidence with most of deuterons detected by the Barrel are beyond the MDM acceptance, the barrel was used mostly for monitoring  elastic scattering.

\subsection{MDM spectrometer}

The MDM spectrometer downstream of the target transports particles scattered at forward angles less than $\theta_{lab}$=$\pm$2$^\circ$ in both the dispersive (x) and non-dispersive (y) planes. Transported particles are detected in the Oxford detector \cite{Pringle1986, Spiridon2016}. 
The Oxford detector consists of three ionization detector zones and four wire proportional counter zones. Energy deposit signals in the two ionization zones in the downstream side are amplified with Micrcromegas plates. The isobutane gas pressure was adjusted so that the incident recoil ions of interest are stopped in the last ionization zone (35 torr for ($^6$Li,$d$) and 70 torr for ($d$,$p$) reactions, respectively). Thus recoil particles were identified by $\Delta$E-E method with high energy-resolution. The resistive wires provided the particle positions and trajectories with a few mm resolution in x-z plane. Thus particles incident on the Oxford detector are identified in mass (A) and atomic number (Z) from the E,  $\Delta$E, and x-position information at the focal plane position, which is approximately located at the second resistive wire.
The charge state (Q) distribution of $^{26}$Mg after passing the target is estimated to be dominated by 12$^+$ \cite{LISE2018}. 
Thus the magnet rigidity was set to accept $^{25,26}$Mg recoils with charge $Q=12$.

\subsection{HPGe array}

\begin{figure}
 \includegraphics[width=8.6cm]{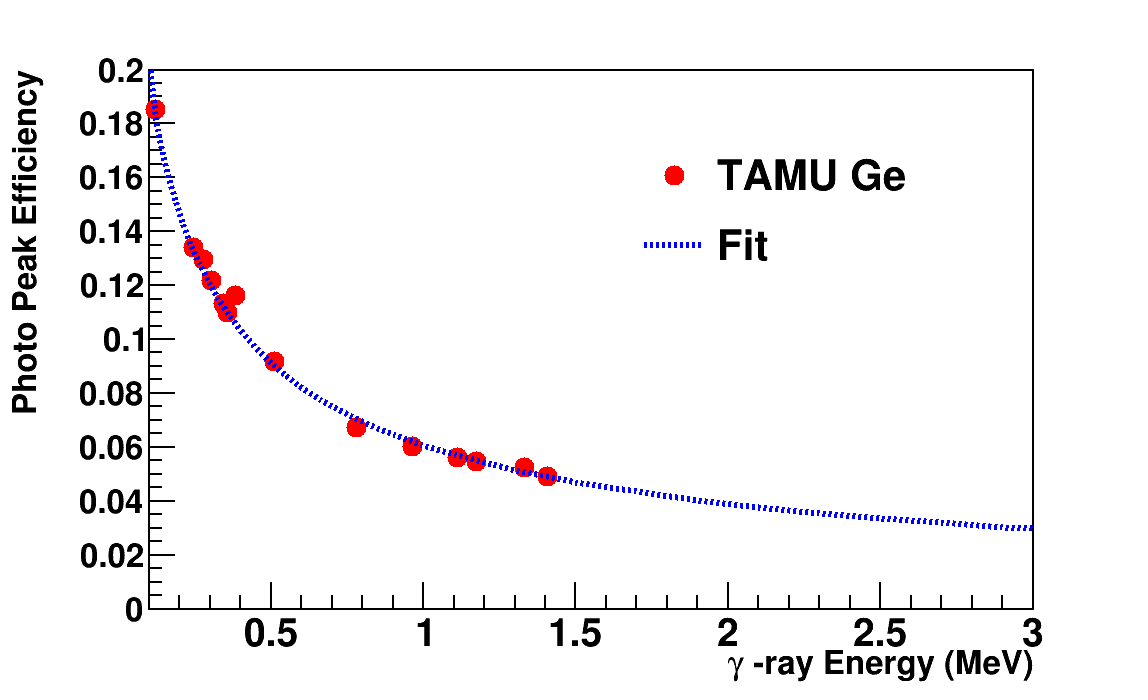}

 \caption{Total photopeak efficiency (sum of efficiency by each clover), together with data obtained using some conventional $\gamma$ sources. 
 \label{fig:Fig2}}
 \end{figure}

Four closely-packed HPGe clovers \cite{Lesher2010} were placed at a distance of about 10 cm from the target position (distance is quoted to  the detector surface). 
Each clover consists of four crystals and is also electrically segmented into three sections toward the beam direction (downstream, middle, upstream). The segmentation information was used to correct for doppler shift of the $\gamma$-rays emitted from beam-like recoils ($\beta \sim 0.1$). The energy resolution achieved for beam recoil $\gamma$-rays is 2--3\% at FWHM for 1 MeV $\gamma$ rays. This is primarily limited by angular resolution in the Doppler correction.
Photopeak efficiency was measured using some conventional $\gamma$-ray sources. Absolute efficiency was determined from the coincidence measurements of $^{60}$Co $\gamma$-rays (1173 and 1332 keV), and $^{22}$Na pair-production $\gamma$-rays (511 keV). The total efficiency curve was then determined with $^{152}$Eu and $^{133}$Ba source measurements, normalizing to the absolute efficiency measurements. Figure~\ref{fig:Fig2} shows the measured photopeak efficiency for this setup, together with a fit to the following function ($E_\gamma >$0.1 MeV): $\Sigma_{i=1}^4 exp(a_i+b_i\times ln(E_\gamma/E_0)+c_i\times ln(E_\gamma/E_0)^2)$ \cite{Knoll2010}, where a fixed reference energy $E_0$ is 1 MeV and the average values of $a_i, b_i, c_i$ for each clover $i$ are $-$4.1979, $-$0.6096, and $-$0.0353, respectively. 

The distance from the chamber surface is about 1 cm. Because of the proximity to the target position, photopeak $\gamma$-ray detection efficiencies achieved in this geometry are 23.5\%, 9\%, and 6\%, and $\sim$1\% (estimated) at 100, 500, 1, and 10 MeV, respectively. Since $\gamma$-ray energies of our interest range from a few hundred keV \cite{Glatz1986, NNDC2018} to 11 MeV \cite{Longland2009}, $\gamma$-rays were simultaneously measured with two shaping amplifiers with different gains: ``low gain" (up to 20 MeV) and ``high gain" (up to 6 MeV). 
The two $\gamma$ spectra were finally merged into one spectrum with $E_\gamma$ = 2.5--3 MeV (depending on crystals) as a cross-over point (i.e., below $E_\gamma$, the high gain spectrum was used).

\subsection{Test measurement with $^{22}$Ne($d,p$)} 
\label{sec:dp}

To test the reliability of our experimental setup, we performed a measurement of $^{22}$Ne($d,p$) reaction, whose cross sections, $J^{\pi}$, and spectroscopic factors are well determined by Refs.~\cite{Lutz1967, Howard1970}.
The beam impinged on a 500 $\mu$g/cm$^2$ CD$_{2}$ target with the intensity of 1 pnA. 
%
The outgoing target-like protons were detected in the Hyball which measured their kinetic energy and angle with respect to the incoming beam. The reconstructed deuteron momenta were used to determine the excitation energy of \nuc{23}{Ne} states using the missing mass technique. Elastically scattered target nuclei were detected in the TIARA Si barrel array and used to continuously monitor the incoming beam rate. 

\begin{figure}
 \includegraphics[width=8.6cm]{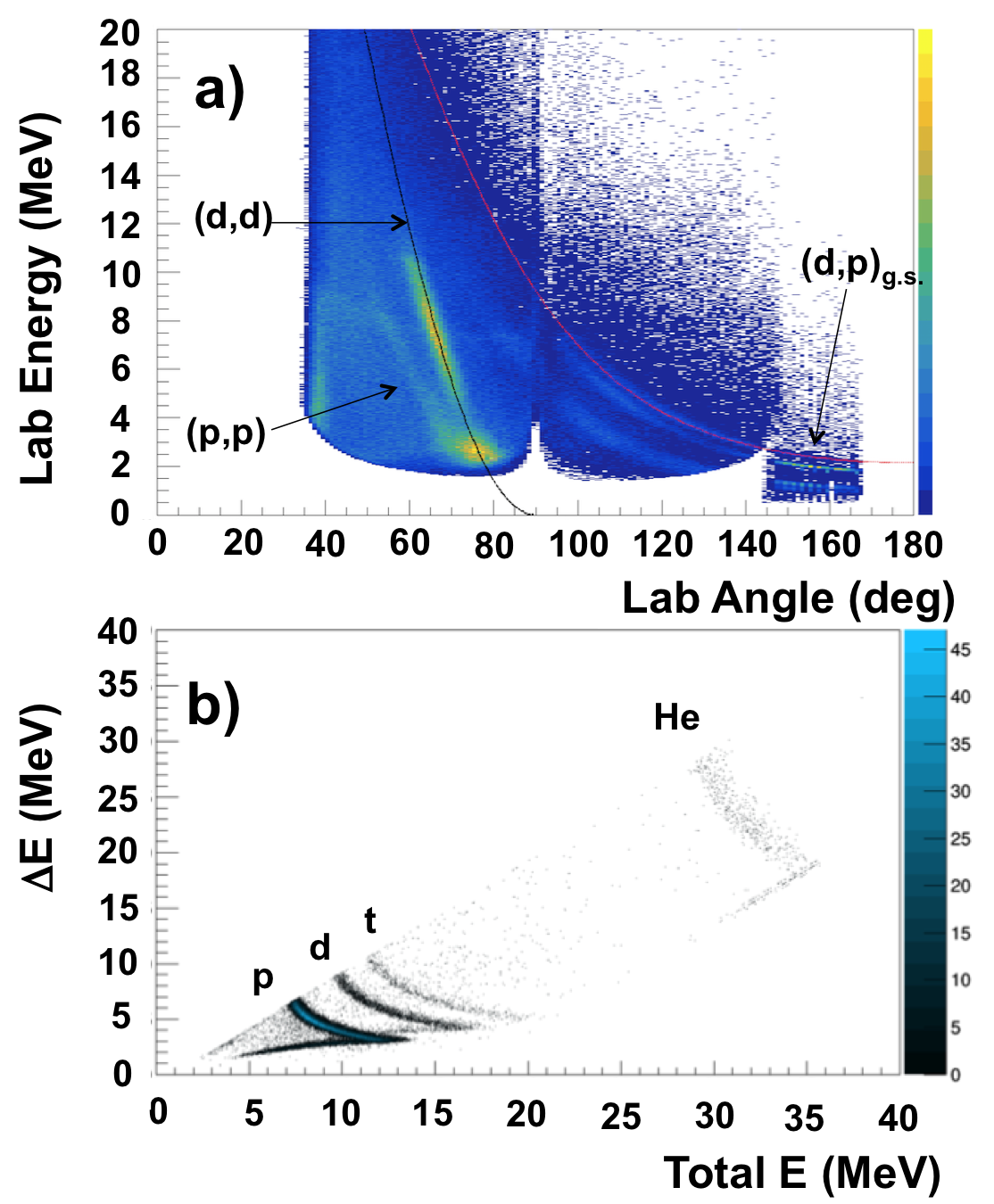} 
 \caption{a) Energy versus scattering angle plot from $^{22}$Ne+CD$_2$. Theoretical elastic ($d,d)$ and ($d,p$) ground state kinematic lines are shown together. Elastic ($p,p)$ line from contaminants in the target. b) $E$-$\Delta E$ plot from a Barrel detector, where protons, deuterons, tritons, and He are observed. 
 \label{fig:Fig3}}
 \end{figure}

Figure~\ref{fig:Fig3} shows an angle versus energy plot, in which theoretically calculated kinematic curves are also drawn. 
For $(d,d)$  elastic scattering events, we can see the expected correlation between the deuteron energy and scattering angles. The location of the elastic scattering kinematic curves was used to determine the position of the incoming beam, by comparing theoretical curves over a range of incident beam positions using a $\chi^2$ minimization technique. 
From this procedure, we obtained a beam position of {x,y,z} = {--2, --1, 4} mm, which was used in later analysis of $(d,p)$ and $({}^{6}\mathrm{Li}, d)$ data as well.
The beam size was also evaluated from the observed elastic scattering kinematic curves, and determined to have a FWHM of 4--5  mm. This agrees with the size determined from the luminescence produced by impinging the beam onto a phosphor-coated viewer plate. 

The elastic scattering cross sections observed in the Barrel detector were compared with optical model calculations to determine the absolute beam + target luminosity normalization.
A variety of available optical potential models \cite{Lutz1967, Howard1970, Daehnick1980, An2006}
were used to calculate the elastic cross sections using the FRESCO code \cite{Thompson1988} and compared with our experimental data. The optical potential by Daehnick \emph{et al}.\ \cite{Daehnick1980} best reproduces the shape of our elastic cross section data and the beam intensity was normalized using this potential. 

\begin{figure}
  \includegraphics[width=8.6cm]{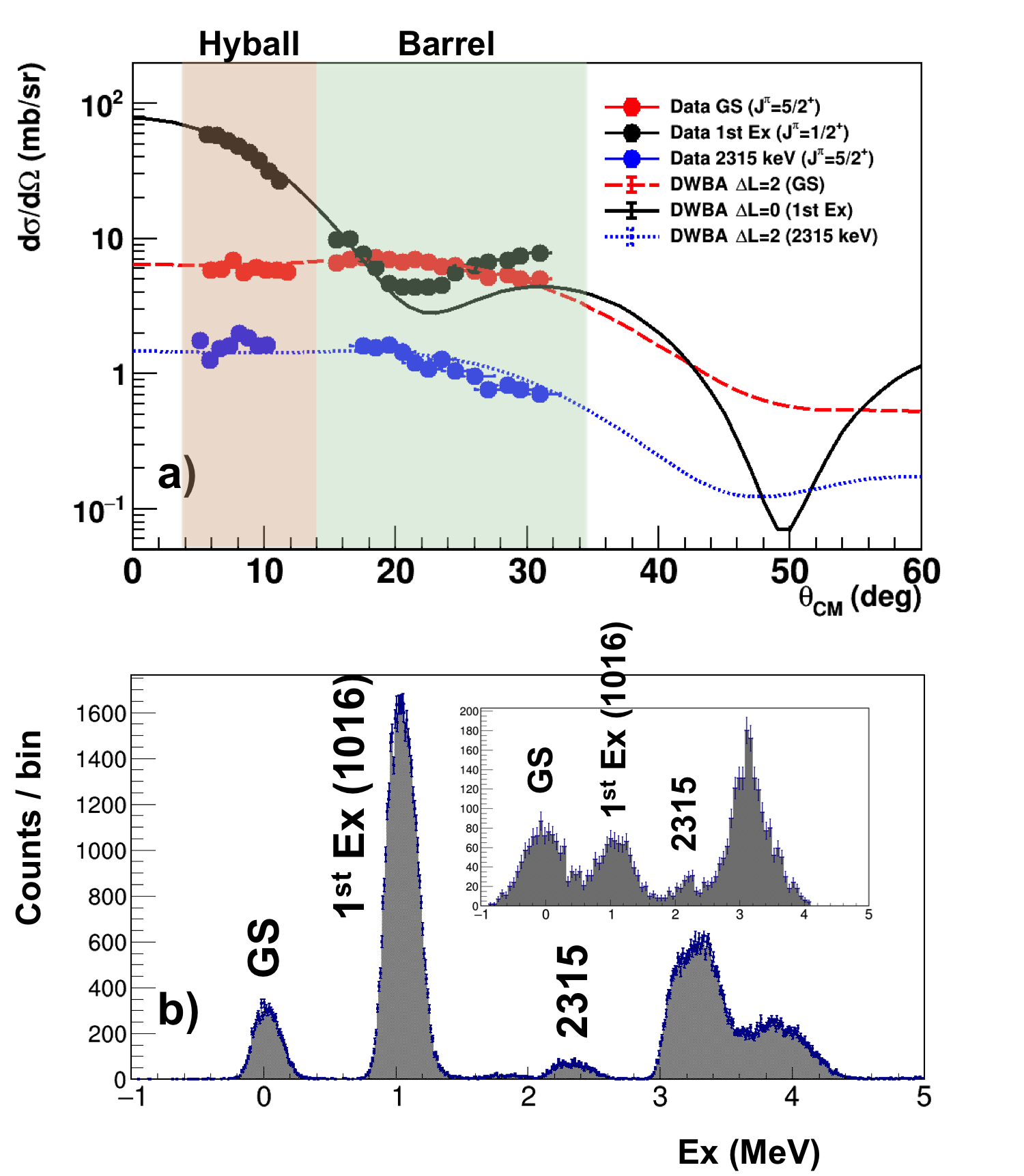}
 \caption{a) Angular differential cross sections of $^{22}$Ne($d,p$) reactions for populating low-lying states of $^{23}$Ne. b) Excitation spectra of ${}^{23}$Ne from the Hyball at $\theta_{CM}$=5--12$^{\circ}$ (whole detector). Inset: Barrel at at $\theta_{CM}$=18--19$^{\circ}$. 
 \label{fig:Fig4}}
 \end{figure}

The $^{22}$Ne($d,p$) excitation spectrum and cross sections for some low-lying states are shown in Figure~\ref{fig:Fig4}, together with DWBA calculations from the TWOFNR code \cite{Twofnr20} using the optical potentials from Refs.\ \cite{Daehnick1980} and ~\cite{Perey1976} for the deuteron and proton channels, respectively, assuming known $J^{\pi}$. 
Our data are well reproduced by the model calculations. Extracted spectroscopic factors are listed in Table~\ref{tab:tab1}. 
These values agree with past measurements, thus confirming that our measurement system works well. 

The ${}^{22}\mathrm{Ne}(d,p)$ data were also used to determine the intrinsic efficiency of the Oxford detector for detecting beam-like recoils.
This was measured to be $80.0 \pm 2.0\%$ by dividing the number of proton~+ $ {}^{23}$Ne coincidence events by the number of proton singles events. For these calculations, the MDM angular acceptance was taken to be $100\%$ since recoils in coincidence with the Hyball protons are emitted within the $\pm 2^\circ$ entrance aperture.
%
The MDM angular acceptance declines with decreasing proton scattering angle, dropping below $100\%$ for events detected in the Barrel. For these events, the coincidence efficiency was estimated using Geant4 with the NPTOOL interface \cite{Matta2016}. This efficiency correction for Barrel events is included in the cross sections shown in Figure~\ref{fig:Fig4}.


The ${}^{22}\mathrm{Ne}(d,p)$ data were also used to test and optimize our particle-$\gamma$ coincidence measurements. 
Because of Doppler broadening, the  $\gamma$-ray energy resolution is strongly affected by the accuracy of the HPGe detector positions. 
Hence the position of each HPGe detector was systematically varied, and the position resulting in the narrowest Doppler-corrected energy resolution was taken for the final analysis.
The photopeak efficiencies for selected $\gamma$-ray transitions observed in coincidence with $(d,p)$ events were also used to confirm the accuracy of our source-determined efficiency curve out to high $\gamma$-ray energies (${\sim} 4$~MeV).
%


\section{$^{22}$Ne + $^{6}$Li Experiment}

For the $^{22}$Ne + $^{6}$Li portion of the experiment, the $^{22}$Ne beam impinged on a 30~$\mu$g/cm$^2$ $^{6}$LiF target (95.0\% \nuc{6}{Li} purity) with a 10 $\mu$g/cm$^2$ carbon backing, with an intensity of 3 pnA. Recoil particle identification plots based on $E$-$\Delta E$ measurements in the Oxford detector, as well a plot of the $^{26}$Mg excitation energy spectrum gated on ${}^{25}$Mg and  ${}^{26}$Mg recoils can be found in Ref.~\cite{Ota2019}. 
Unlike the $(d,p)$ measurements, the kinematic locus of ${}^{22}\mathrm{Ne} + {}^{6}\mathrm{Li}$ elastic scattering events is contaminated by background from elastic scattering on the  fluorine and carbon contained in the target. Therefore, we focused on determining relative $\alpha$-particle strengths, normalized to the well-constrained state at $E_x$=11.32 MeV (see Ref.~\cite{Ota2019}). Note the Barrel was not used for the $^{22}$Ne + $^{6}$Li analysis due to its poor excitation energy resolution.

 \begin{table}
\caption{\label{tab:tab1} Obtained spectroscopic factors for some low-lying states of $^{23}$Ne observed in ${}^{22}\mathrm{Ne}(d,p)$, compared with values from past measurements. }
\begin{ruledtabular}
\begin{tabular}{ccccc}
$E_x$ (keV) & $J^\pi$ & $S$ (present) & $S$ (Ref.~\cite{Howard1970}) & $S$ (Ref.~\cite{Lutz1967}) \\
\hline
$GS$  
         & $2^+$ & $0.25\pm0.05$ & 0.22 & 0.24\\
$1016$  
         & $0^+$ & $0.58\pm0.12$ & 0.70 & 0.40\\
$2315$  
         & $2^+$ & $0.022\pm0.05$ & 0.05$\pm$0.01 & 0.07\\
\end{tabular}
\end{ruledtabular}
\end{table}

%

\subsection{Recoil energy and focal plane position}

\begin{figure*}[!ht]
        \centering
          \includegraphics[width=18cm]{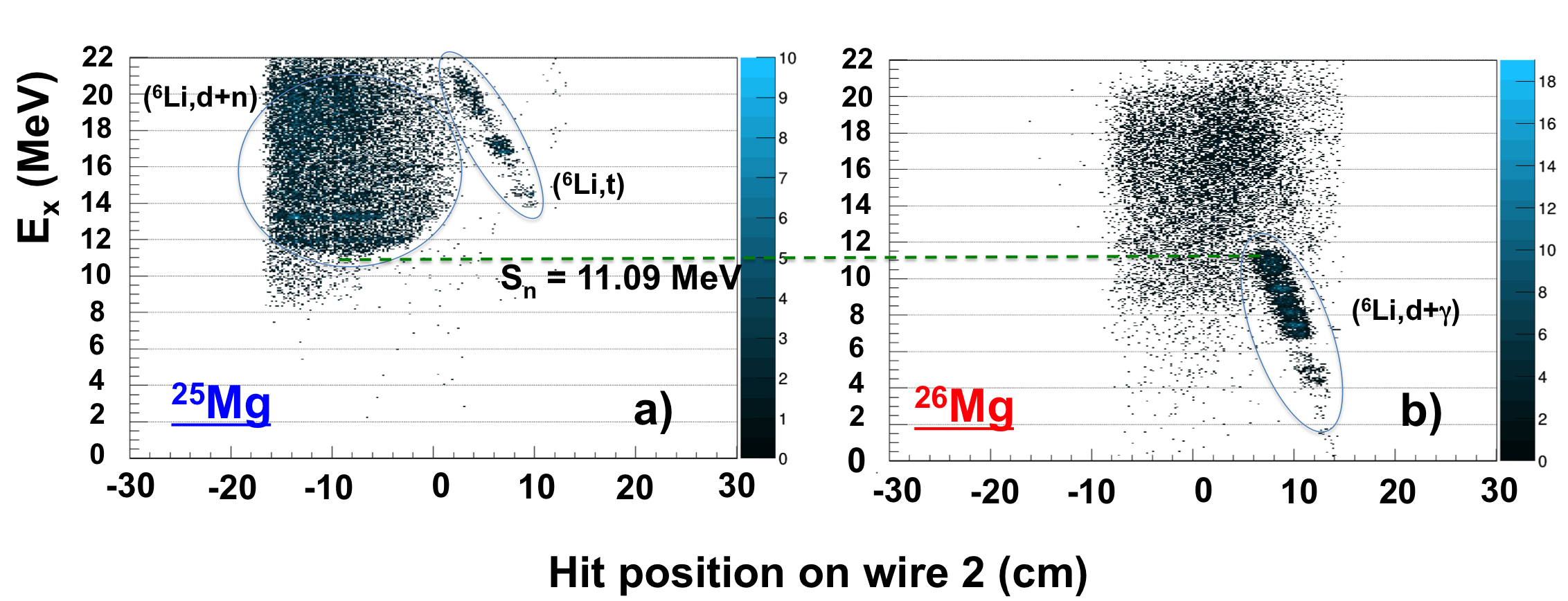}          
        \caption{$E_x$ versus hit position on the focal plane in the Oxford detector. All excitation energies are constructed from Hyball-detected light particle momenta, assuming the $^{22}$Ne($^6$Li,$d$) reaction. Left panel: gated on $^{26}$Mg recoil. Right panel: gated on $^{25}$Mg recoil. Clear correlations from the binary reactions ($^6$Li,$d$) and ($^6$Li,$t$) can be observed. In the $^{25}$Mg recoils, ($^6$Li,$d$) kinematic lines are spread in x-direction due to neutron evaporation. Transition from $^{26}$Mg to $^{25}$Mg is clearly occurring at the neutron separation energy of $^{26}$Mg (11.09 MeV). 
        }\label{fig:Fig5}
    \hfill{}
\end{figure*}

Figure~\ref{fig:Fig5} shows the correlation of the recoil particle positions at the focal plane and $^{26}$Mg excitation energies calculated from the Hyball signals assuming the $^{22}$Ne($^6$Li,$d$)$^{26}$Mg reaction. 
Clear correlations are seen in both figures, which are gated on $^{25,26}$Mg recoils, respectively. These correlated loci correspond to two separate binary reaction mechanisms as explained below.
Since the Hyball detector does not have particle identification capability, 
this correlation is essential to separate the two binary reactions observed in this study.
The observed correlations occur because the $^{25,26}$Mg recoils have lower kinematic energies when they are more highly excited in the binary reactions. Thus recoils with increasing excitation energy move towards the lower-rigidity (decreased $x$ position) side of the focal plane.
It is also noticeable that the correlation disappears from $^{26}$Mg events when the excitation energy becomes greater than the neutron separation energy (11.09 MeV). 
Instead, the excited ${}^{26}$Mg decays into ${}^{25}\mathrm{Mg} + n$. This process gives a substantial momentum kick to ${}^{25}$Mg following the neutron decay, which effectively destroys the kinematic correlations especially for large neutron decay energies (e.g., $E_x$ $>$ 12 MeV).
%

\begin{figure}
 \includegraphics[width=9cm]{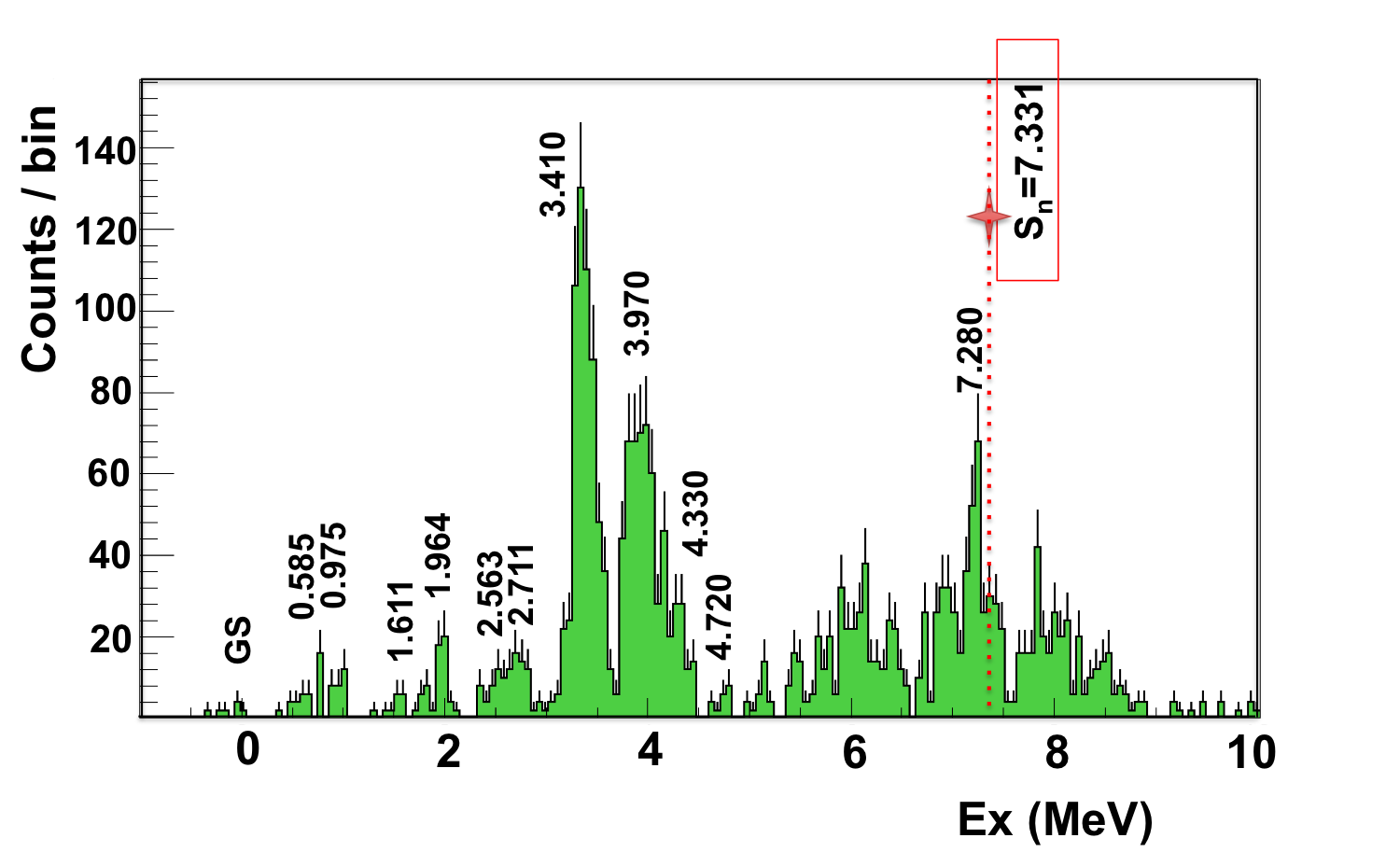}
 \caption{$^{25}$Mg excitation energy spectrum measured from the $^6\mathrm{Li}({}^{22}\mathrm{Ne},t){}^{25}$Mg reaction at $\theta_{CM}$=7$^{\circ}$--14$^{\circ}$. 
 \label{fig:Fig6}}
 \end{figure}

If only $^{22}$Ne($^6$Li,$d$)$^{26}$Mg reactions were present, it would be expected that the $x$-$E_x$ plot gated on ${}^{25}$Mg events would display no kinematic correlation, as all events would arise from the $^{22}$Ne($^6$Li,$d$)$^{26}\mathrm{Mg} \rightarrow {}^{25}\mathrm{Mg} + n$ process described above. However, it is clear from Figure~\ref{fig:Fig5}(b) that a correlated locus occurs in the high excitation energy and the large positive $x$-position side of the plot. 
This correlation occurs because these events arise from the $^{22}$Ne($^6$Li,$t$)$^{25}$Mg reaction. The association of these events with the $({}^{6}\mathrm{Li},t)$ reaction channel is confirmed by calculating the excitation energy assuming the $({}^{6}\mathrm{Li},t)$ reaction mechanism. When calculated as such, the excitation energy spectrum starts from 0 MeV as would be expected (see Figure~\ref{fig:Fig6}).  This reaction mechanism is further discussed in Section~\ref{sec:Li6t}. 

In the figures, background events can be seen behind the correlated loci of binary reactions.
These events are mostly due to protons arising from compound nuclear reactions.
We confirmed this by constructing the same plots using Barrel-MDM coincidence events. In the barrel, light-particle identification is possible using the $\Delta E$-$E$ method (see \ref{fig:Fig3}(b)), and the observed uncorrelated events were thus identified to be protons.
%
In the binary-reaction analysis, the shape of the proton background was estimated from the uncorrelated events located near the binary events in Figure~\ref{fig:Fig5}. These background contributions composed ${\sim} 5$--$10\%$ of the total events, and were subtracted in bulk in the final excitation energy spectrum.



 \begin{figure*}[!ht]
        \centering
 \includegraphics[width=18cm]{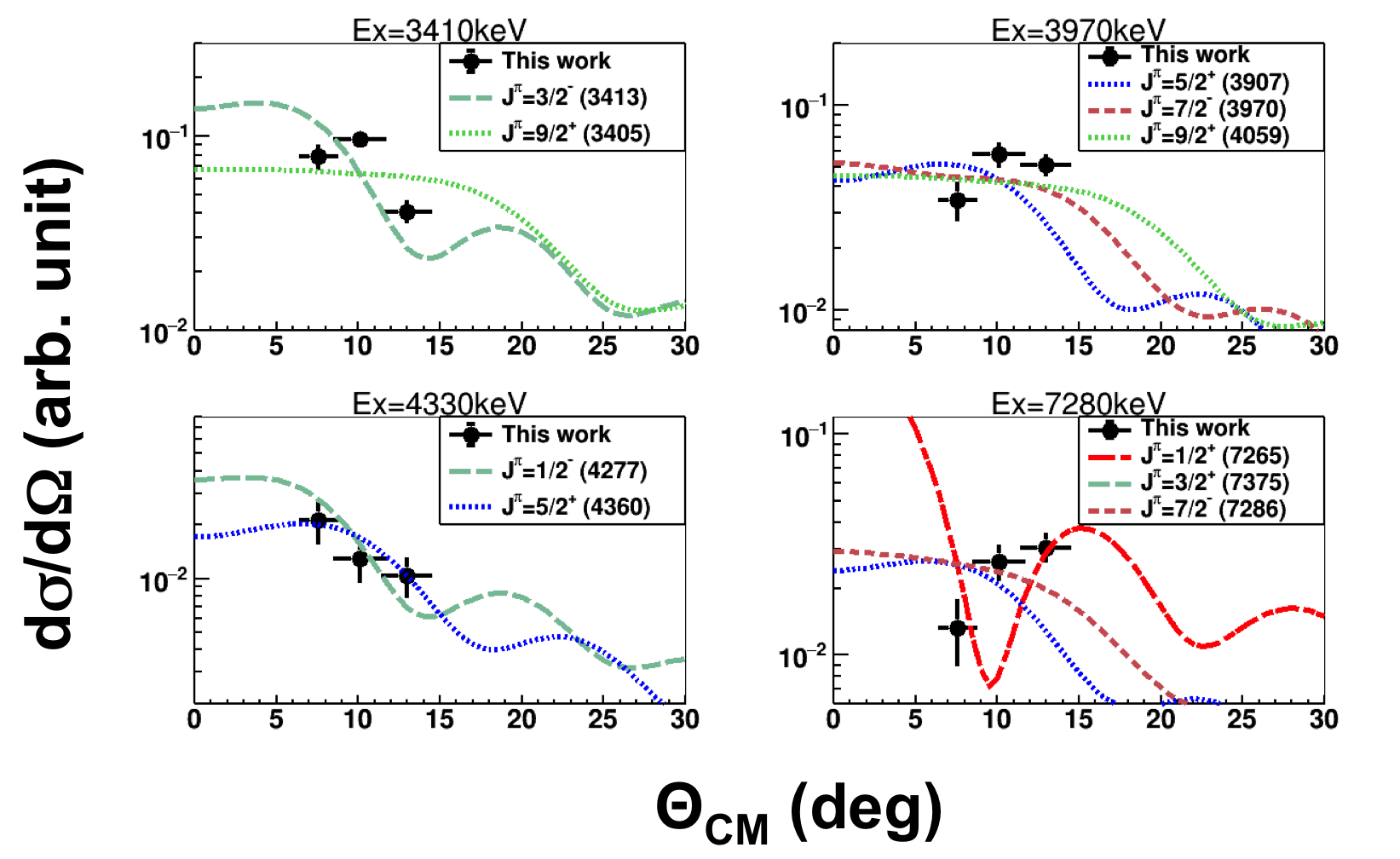}
 \caption{ Angular differential cross sections of  $^{22}$Ne($^6$Li,$t$) reaction for populating various states of $^{25}$Mg, compared with DWBA calculations.  
 \label{fig:Fig7}}
    \hfill{}
\end{figure*}

%
%

\subsection{$^{22}$Ne($^{6}$Li,$t$)$^{25}$Mg reaction}
\label{sec:Li6t}



 We analyzed the $^{22}$Ne($^6$Li,$t$)$^{25}$Mg  data to identify the states in ${}^{25}$Mg that are strongly populated in this reaction and determine their 
 spectroscopic factors. To date, there are no published data on the $^{22}$Ne($^6$Li,$t$)$^{25}$Mg reaction and hence this represents the first analysis of states in ${}^{25}$Mg populated through this reaction mechanism.
However, the ($^6$Li,$t$) reaction mechanism has been studied on a variety of light-to-medium mass targets including $^{13}$C ~\cite{Bassani1969, Bassani1971,Avila2018}, $^{16}$O ~\cite{Panagiotou1972, Garrett1972, Martz1979}, $^{24}$Mg, $^{28}$Si, $^{40}$Ca, $^{54,56}$Fe, and $^{58}$Ni ~\cite{Lindgren1974, Woods1978},  as a potential method for spectroscopy. Most of these studies were carried out  in the 1970s or earlier except for the recent $^{13}$C($^6$Li,$t$) study of Ref.~\cite{Avila2018}. 
Despite the large binding energy (15.8 MeV), appreciable $t$+$^3$He clustering in the ground state of $^6$Li is reported in those past studies.

A few potential direct reaction mechanisms were proposed in the earlier  ($^6$Li,$t$)  studies. One is that the proton pair goes into the lowest accessible orbits, while the odd neutron is transferred into single-particle states (in the present experiment, this would populate a similar set of ${}^{25}$Mg states as $^{24}${Mg}($d,p$)$ {}^{25}$Mg) \cite{Woods1978}. 
Another possibility is that one proton is transferred to the lowest-available orbital and the remaining proton-neutron pair is transferred as in the ($\alpha$,$d$) reaction (analogous to $^{23}${Na}($\alpha,d$)$ {}^{25}$Mg in the present case) \cite{Bassani1971}.  Finally, an analogous mechanism to direct ($\alpha,n$) reactions with transfer of a $^{3}$He cluster ($^{22}${Ne}($\alpha,n$)$ {}^{25}$Mg at present) has also been proposed  ~\cite{Woods1978}. 
However, the reaction's selectivity in populating the specific levels observed in the past experiments ~\cite{Bassani1969, Bassani1971, Panagiotou1972, Lindgren1974, Woods1978} are not universally explained by any of the above mechanisms. 

The background-subtracted $^{25}$Mg excitation spectrum from the present experiment is shown in Figure~\ref{fig:Fig6}. 
A number of narrow small peaks and some large peaks are evident, indicating the strongly selective population of discrete states through a direct transfer mechanism.
This is consistent with existing  ($^6$Li,$t$) studies. 
%
Due to the low statistics, we were unable to perform a particle-$\gamma$ coincidence analysis to unambiguously identify the populated states through their $\gamma$-ray transitions.
%
Instead, we made a tentative assignment of the populated states based purely on excitation energies observed from $({}^{6}\mathrm{Li},t)$ (the excitation energy uncertainty is $\pm$20 keV). Due to the energy resolution of the present setup ($\sim$200 keV), 
it is possible that  the observed peaks may contain transitions to two or more states. Hence, for each peak we have performed a separate analysis of all candidate states that overlap in energy.

The strongest peaks in the  observed spectrum appear at $E_x$=3410 ($J^{\pi}$=3/2$^-$), 3973 (7/2$^-$), 4277 (1/2$^-$), and 7280 (7/2$^-$) keV, respectively.  
Notably, these peaks all appear at energies that are close to known negative parity states, only a handful of which exist, especially in the low energy range ($E_x$$<$5 MeV) \cite{NNDC2018}. 
If these peaks do indeed correspond to the negative parity candidate states, this suggests a possible strong contribution from the $fp$ shell. 
The $E_x$=3410, 3973, 4277 keV states are rotational members of the $K$=1/2$^{-}$ band ([$Nn_z\Lambda\Omega$]=[330 1/2]), which is obtained from one-particle excitation in the $fp$-shell ~\cite{Ropke1974, Heidinger1991, Hamamoto2012}. 
It is therefore reasonable to deduce that the mechanism is similar to transfer of a neutron into the $fp$ shell in $^{24}$Mg($d,p)$. 
The spectra obtained by $^{24}$Mg($d,p\gamma)$ reactions are indeed similar to our spectra to some degree, e.g.\ showing highly populated $E_x$=3410, 3973, 4277, and $\sim$7200 keV states ~\cite{Heidinger1991, Meurders1975, Hinds1957}. 

Concerning the other two proposed transfer mechanisms, there are currently no $^{23}$Na($\alpha,d$) reaction data available for comparison, and while some relevant $^{22}$Ne($\alpha,n\gamma$) data do exist~\cite{Butler1975, Christiansson1976, Heidinger1991}, the associated publications do not show excitation spectra of $^{25}$Mg or report specific selectivity of the reaction. 
As a result comparisons with ($\alpha,d$) or  ($\alpha,n$) spectra are not currently possible.

Another point of interest is that our ($^6$Li,$t$)$^{25}$Mg spectra continues beyond the neutron separation energy ($S_n$=7331 keV) up to $E_x$=8.5--9 MeV, without neutron decaying into $^{24}$Mg (note the spectrum is gated on $^{25}$Mg). 
This indicates high spin states,  $I \ge ({11}/{2})\hbar$  are populated since these states would have a suppressed neutron decay probability resulting from the large  centrifugal barrier.
The enhancement of high-spin states is to be expected due to the highly negative $Q$-value of the $({}^{6}\mathrm{Li},t)$ reaction ($-15.8$~MeV). However, we also note that similar
high spin states were observed by the $^{24}$Mg($d,p\gamma$) reaction
 ~\cite{Heidinger1991}. 

Figure~\ref{fig:Fig7} shows the angular distribution of $E_x$=3410, 3973, 4330, and 7280 keV peaks. 
DWBA calculations were made in the same manner as Ref.~\cite{Ota2019} using the optical potential parameters \cite{Li2007} in Table~\ref{tab:tab3}. The calculations were compared with the data to extract spectroscopic factors given by 
$S_{({}^{6}\mathrm{Li},t)} = (\mathrm{d}\sigma/\mathrm{d}\Omega)_{exp}/(\mathrm{d}\sigma/\mathrm{d}\Omega)_{{DWBA}}$, along with spin-parity assignments. The results of this analysis are shown in Table~\ref{tab:tab2}.
The cross sections and spectroscopic factors were extracted using a similar ${}^{22}\mathrm{Ne} + {}^{6}\mathrm{Li}$  normalization technique to the ${}^{22}\mathrm{Ne}(d,p)$ data (Section~\ref{sec:dp}), i.e.\ normalizing to elastic scattering cross sections in the Barrel detector.  
While this technique produced cross sections that are in the ballpark of expected 
($^6$Li,$t$) cross sections, the elastic-scattering spectra unfortunately contained unknown levels of contamination from F and C components of the target.
As a result, the cross sections and spectroscopic factors should be cautiously treated as being arbitrarily normalized. However, comparisons of relative spectroscopic factors between states are expected to be robust.

%
%

In the following paragraph, we discuss some of the extracted features of the most strongly observed peaks in the spectrum.
Since these peaks may contain two or more unresolved states, we have used the shapes of the measured angular distributions to determine the most likely constituents, when possible. Additionally, we have used the published $^{24}$Mg($d,p$) data as a tentative guide to understanding the spectrum. This follows the arguments given previously  that the ($^6$Li,$t$) reaction may populate a similar set of states as $^{24}$Mg($d,p$). A similar analysis was performed for the less strongly populated peaks in the spectrum, with results reported in Table~\ref{tab:tab2}.

The $E_x$=3410 keV peak is likely to be dominated by the 3413 keV ($J^{\pi}$=3/2$^-$) state, as the calculated  angular distribution is a much better match to the data.
This is consistent with the $^{24}$Mg($d,p$) experiment, which also shows a much stronger transition to the 3414 keV state, compared to the 3405 keV state ~\cite{Meurders1975}. 
The $E_x$=3970 keV peak may come from multiple candidate states---3907 keV ($J^{\pi}$=5/2$^+$), 3970 keV ($J^{\pi}$=7/2$^-$), or 4059 keV ($J^{\pi}$=9/2$^+$). All of these candidate states have calculated angular distributions consistent with the data.
For the $E_x$=4277 keV peak, both the 4277 keV ($J^{\pi}$=1/2$^-$) and 4360 keV ($J^{\pi}$=5/2$^+$) keV states are possibilities,  having angular distributions consistent with the data.  
These two peaks, however, are also likely to be dominated by one single state, (3970 and 4277 keV, respectively), again following the $^{24}$Mg($d,p$) experiment. 
The $E_x$=7280 keV peak likely comes from the 7265 keV state ($J^{\pi}$=1/2$^+$).


Despite the difficulty of interpreting the present spectrum, the selectivity observed in detail in the ($^6$Li,$t$) reaction indicates that it may be a useful tool for future nuclear structure studies, e.g., as an experimental alternative to use of ($d,p$) reaction, including experiments utilizing unstable beams in inverse kinematics. The disadvantage is that the theoretical interpretation and the direct reaction mechanism are significantly more complex.
 

 \begin{table*}
\caption{\label{tab:tab2} $^{25}$Mg states populated in the present ($^6$Li,$t$) experiment. Excitation energies and spin-parities are adopted from \cite{NNDC2018}. Relative spectroscopic factors for each state, which were extracted from the cross sections obtained by normalizing to the estimated beam intensity, are also listed. 
}
\begin{ruledtabular}
\begin{tabular}{ccc}
$E_x$ (keV) & $J^\pi$ & $C^2S_{({}^{6}\mathrm{Li},t)}$ \\
\hline
$GS$  
         & $5/2^+$ & $0.021(9)$\\
$585$  
         & $1/2^+$ & $0.036(16)$\\
$974$  
         & $3/2^+$ & $0.114(23)$\\
$1611$  
         & $7/2^+$ & $0.075(25)$\\
$1964$  
         & $5/2^+$ & $0.088(16)$\\
$2563$  
         & $1/2^+$ & $0.157(33)$\\
$2737, 2801$  
         & $7/2^+,3/2^+$ & $0.202(36),0.122(22)$\\
$3405, 3413$  
         & $9/2^+,3/2^-$ & $0.45(7),0.33(5)$\\
$3907, 3970, 4059$  
         & $5/2^+,7/2^-2,9/2^+$ & $0.16(2),0.10(2),0.26(4)$\\
$4277, 4359$  
         & $1/2^-, 3/2^+$ & $0.155(23), 0.09(2)$\\
$4711, 4722$  
         & $9/2^+, (3/2^+, 5/2^+)$ & $0.031(10), 0.030(10), 0.020(7)$\\
$7265,7286, 7375$  
         & $1/2^+,7/2^-,3/2^+$ & $0.70(10),0.036(5),0.09(1)$\\
\end{tabular}
\end{ruledtabular}
\end{table*}


\begin{table*}
\caption{\label{tab:tab3}Optical parameters used in FRESCO for DWBA analysis of $^{22}$Ne($^6$Li,$d$)$^{26}$Mg \cite{Ota2019} and $^{22}$Ne($^6$Li,$t$)$^{25}$Mg} \cite{Li2007}. All radii except those for the $\alpha$ + $d$ channel are given such that $R_x =r_x A^{1/3}$. 
For the $\alpha+d$ channel, $R_x =r_x$.
\begin{ruledtabular}
\begin{tabular}{cccccccccccccc}
 Channel & $r_c$ & $V_r$ & $r_r$ & $a_r$ & $W_i$ & $r_i$ & $a_i$ & $W_D$ & $r_D$ & $a_D$ & $V_{so}$ & $r_{so}$ & $a_{so}$\\
  & (fm) & (MeV) & (fm) & (fm) & (MeV) & (fm) & (fm) & (MeV) & (fm) & (fm) & (MeV) & (fm) & (fm)\\
\hline
 $^{22}$Ne+$^6$Li & 1.30 & 117.04 & 1.80 & 0.40 & 48.60 & 1.99 & 0.62\\
 $^{26}$Mg+$d$ & 1.30 & 79.07 & 1.17 & 0.79 & 2.99 & 1.33 & 0.74 & 10.51 & 1.33 & 0.74 & 5.88 & 1.07 & 0.66\\
 $\alpha$+$d$ & 1.90 & {\footnotemark[1]} & 1.90 & 0.65 & & & & &\\
  Final State & 1.40 & {\footnotemark[2]} & 1.40 & 0.70 & & & &\\ 
   $^{25}$Mg+$t$ & 1.42 & 149.57 & 1.07 & 0.74 & 9.65 & 1.26 & 1.18 & 31.95 & 1.09 & 0.85 &1.90 & 0.51 & 0.20\\
    $^3$He+$t$ & 1.25 & {\footnotemark[1]} & 1.25 & 0.65 & & & & &\\
  Final State & 1.40 & {\footnotemark[3]} & 1.40 & 0.70 & & & &\\ 
\end{tabular}
\end{ruledtabular}
\footnotetext[1]{Adjusted to give the correct $^6$Li binding energy}
\footnotetext[2]{Adjusted to give the correct final state binding/resonance energy}
\footnotetext[3]{Adjusted to give the correct final state binding energy}
\end{table*}

\section{$^{22}$Ne($^{6}$Li,$d$)$^{26}$Mg reaction}
\label{sec:Li6d}

\begin{figure*}
  \begin{center}
 \includegraphics[width=\textwidth]{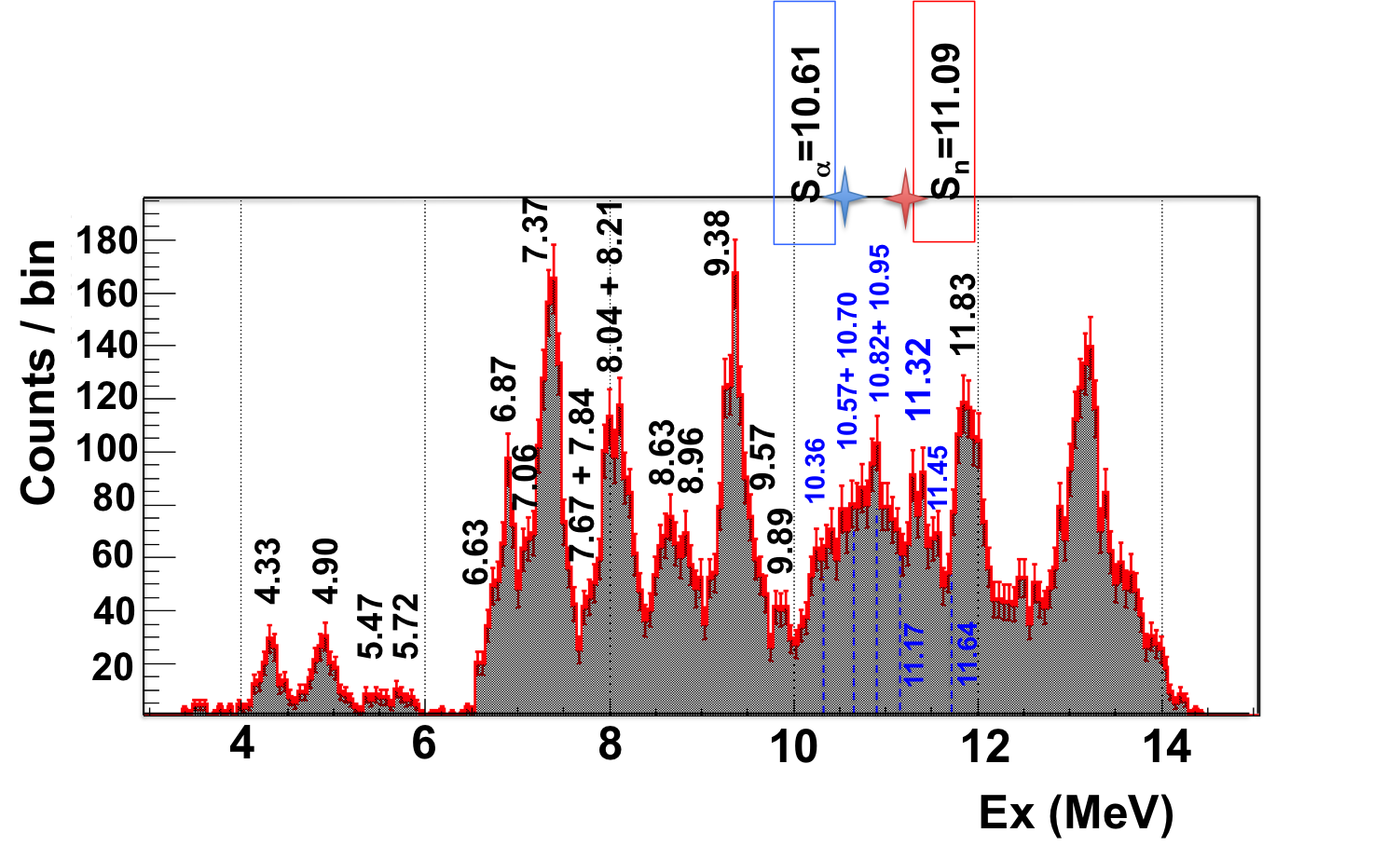} 
 \caption{$^{26}$Mg excitation energy spectrum measured from the $^6\mathrm{Li}({}^{22}\mathrm{Ne},d){}^{26}$Mg reaction at $\theta_{CM}$=7$^{\circ}$--14$^{\circ}$. All states considered in the present data analysis are labeled in the figure. These states are mostly determined with the help of the coincident $\gamma$ rays and past ($^6$Li,$d$) experiments. 
 \label{fig:Fig8}}
 \end{center}
 \end{figure*}

Figure~\ref{fig:Fig8} shows the $^{26}$Mg excitation spectrum obtained from the ${}^{22}\mathrm{Ne}({}^{6}\mathrm{Li},d){}^{26}\mathrm{Mg}$ reaction in the present experiment. 
A number of resolved and unresolved peaks are evident.
In the following sections, the states populated in the reaction are identified and analyzed using the angular distributions of the deuterons, along with the coincident $\gamma$~rays detected in the HPGe detectors. The assumed states thus identified are annotated in Figure~\ref{fig:Fig8}.
%
 We also compare our results with those of past (normal kinematics) studies: Refs.~\cite{Anantaraman1977, Giesen1993, Ugalde2007, Talwar2016}, which utilized $^6$Li beams with energies of 5.3 \cite{Anantaraman1977}, 5.3 \cite{Giesen1993}, 5.0 \cite{Ugalde2007}, and 13.5 MeV/nucleon \cite{Talwar2016}, respectively.
The previous studies varied in the range of excitation energies to which they were sensitive: 0--9 MeV  (but with peaks identified and analyzed only up to 5.7~MeV) \cite{Anantaraman1977}, 9.3--12 MeV \cite{Giesen1993}, 9--11 MeV \cite{Ugalde2007}, and 7.3--11.4 MeV \cite{Talwar2016}.  The present experiment is sensitive to $E_x > 4$~MeV. Below $E_x = 4$~MeV, the ejected deuterons punch through the Hyball, and $^{26}$Mg recoils are also outside of the Oxford detector acceptance window.



\begin{figure}
 \includegraphics[width=9cm]{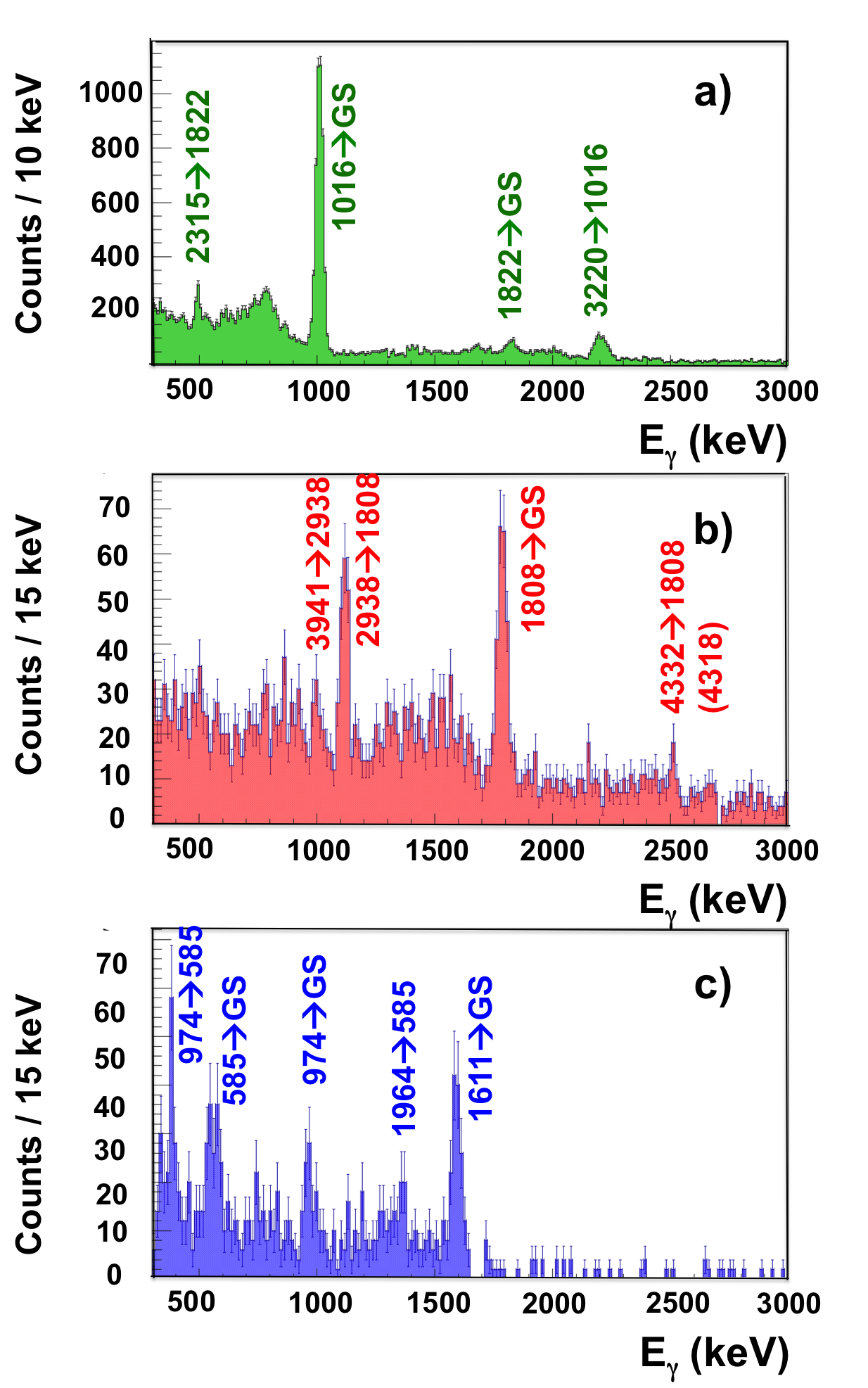} 
 \caption{Panel (a): $\gamma$-ray spectrum from $^{22}$Ne($d,p\gamma$)$^{23}$Ne reaction in coincidence with $E_x$=GS to 5 MeV from the proton excitation-energy spectrum. Panel (b): $\gamma$-ray spectrum from the $^{22}$Ne($^6$Li,$d\gamma$)$^{26}$Mg reaction in coincidence with $E_x$=4--11.5 MeV. Panel (c): $\gamma$-ray spectrum from the $^{22}$Ne($^6$Li,$dn\gamma$)$^{25}$Mg reaction in coincidence with  $E_x$ ($^{26}$Mg) =11.0--14.0 MeV} deuterons. 
 \label{fig:Fig9a}
 \end{figure}

Figure~\ref{fig:Fig9a} shows $\gamma$-ray spectra in coincidence with $^{25,26}$Mg$+$$d$, as well as the spectrum of $\gamma$-rays in coincidence with ${}^{23}\mathrm{Ne} + d$ from the ${}^{22}\mathrm{Ne}(d,p){}^{23}\mathrm{Ne}$ measurement  (Section~\ref{sec:dp}). The high statistics of the latter spectrum make it a useful guide in interpreting the $\gamma$~rays from the ${}^{22}\mathrm{Ne}({}^{6}\mathrm{Li},d\gamma)$ measurement. Before discussing the $\gamma$-ray spectra in coincidence with specific excitation energy regions, we briefly discuss some features of the total spectrum here.
%

In the ${}^{26}$Mg coincidence data, two large peaks (first excited state: 1808$\rightarrow$GS) and (second excited state: 2938$\rightarrow$1808) are evident. This is to be expected as these transitions are fed by a large number of higher-lying states.
Additionally, small peaks (4318/4332/4350$\rightarrow$1808) and (3941$\rightarrow$2938) are present.
These smaller peaks are more easily observed when we gate on a specific excitation energy from the deuteron spectrum, as discussed in the subsequent sections. 
In addition to the prominent peaks, we also observe a background from Compton scattering (our HPGe detectors are not Compton-suppressed). To better understand this, we used the high-statistics $^{22}$Ne($d,p\gamma$) data (Figure~\ref{fig:Fig9a}a) to estimate the ratio of events in the well-separated 1016 and 2203 keV photopeaks to their respective Compton edges. This analysis suggests a  $\sim$20--25\% ratio for $1$--2 MeV $\gamma$-rays.

In the bottom panel of Figure~\ref{fig:Fig9a}, we show the $\gamma$-ray spectrum in coincidence with ${}^{25}$Mg recoils and deuterons in the 11--14 MeV excitation energy range of $^{26}$Mg. The observed $\gamma$~rays correspond to transitions in ${}^{25}$Mg following neutron evaporation of states in $^{26}$Mg.
We observe 384, 585, 974, 1611 keV $\gamma$-ray transitions from the first, second, and third excited states of  $^{25}$Mg, suggesting that states in this region have a significant neutron decay branch to low-lying excited ${}^{25}$Mg states. This is consistent with Ref.~\cite{Wolke1989}, which reported significant $(\alpha, n_1)$ and $(\alpha, n_2)$ strengths for a number of states observed in a direct ${}^{22}\mathrm{Ne}(\alpha,n\gamma){}^{25}\mathrm{Mg}$ experiment. 
Since we do not measure neutrons, it is not realistic to estimate the populated ${}^{26}$Mg states from the $\gamma$-ray transitions alone. However, limited information about the decay properties of ${}^{26}$Mg states in this region can still be obtained from the coincident $\gamma$-ray analysis, as discussed in Section~\ref{ss:Mg25CoincGamma}.

\begin{figure*}
        \centering
          \includegraphics[width=18cm]{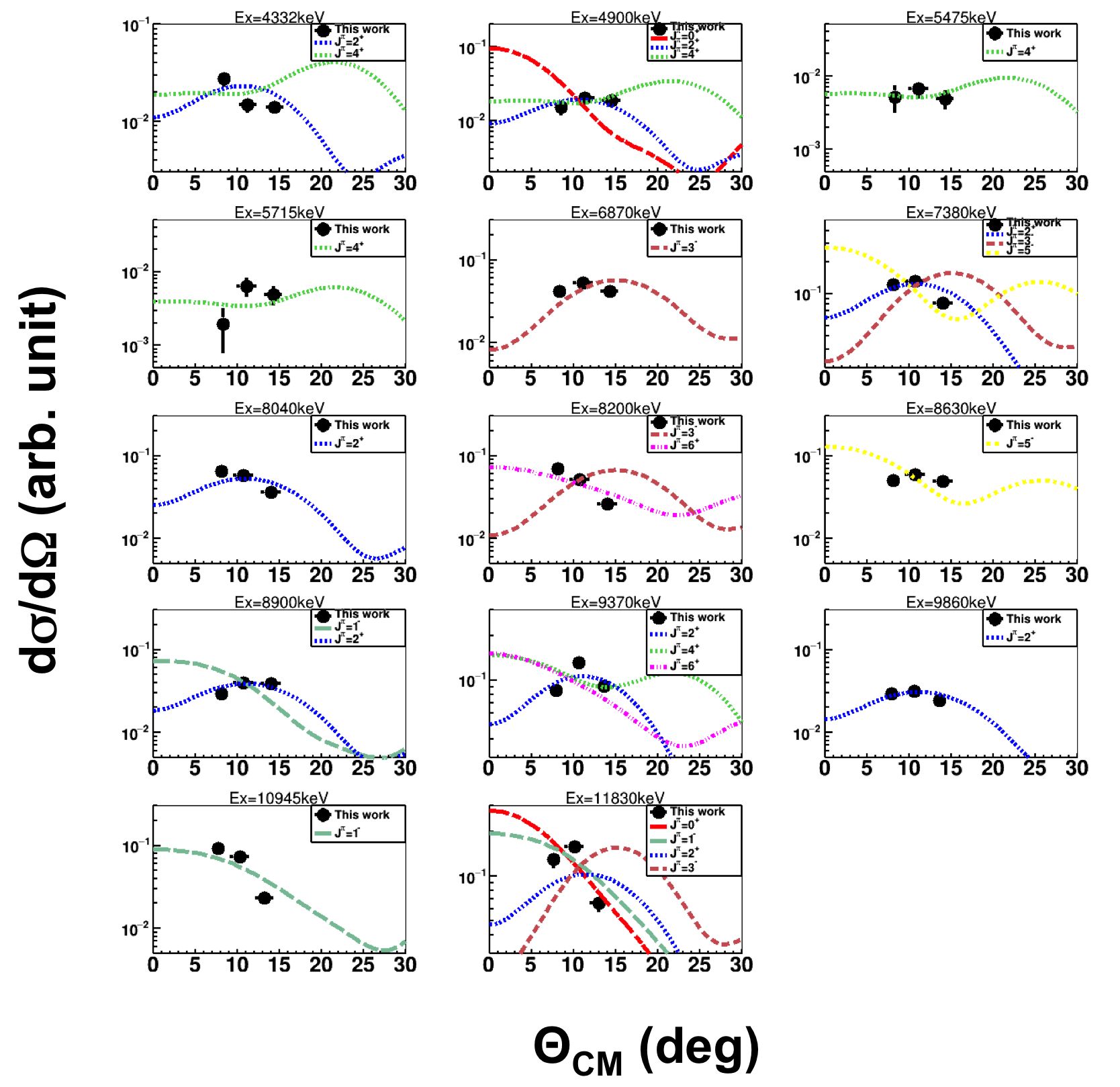}          
        \caption{Angular differential cross sections of $^{22}$Ne($^6$Li,$d$) reactions for populating various states of $^{26}$Mg, compared with DWBA calculations.    
        }\label{fig:Fig10}
\end{figure*}

After identifying the set of populated states, using particle-$\gamma$ coincidences with the help of past ($^6$Li,$d$) data, 
we performed a multiple Gaussian fit to peaks in the deuteron-reconstructed excitation energy spectrum. These fits were then used to determine cross sections and angular distributions of a given peak.
For the fits, the centroid of each Gaussian was taken to be the literature energy of the corresponding state, and the width of each Gaussian was set according to the experimental energy resolution (FWHM 200-250 keV, depending on excitation energy). Amplitudes were allowed to freely vary, and the resulting Gaussian integrals were used to determine a differential cross section for each state. 
The resulting angular distributions for each of the noticable peaks in the deuteron spectrum are displayed in Figure~\ref{fig:Fig10}. Using these distributions,  spins and relative spectroscopic factors were calculated, as presented in Table~\ref{tab:tab4} and discussed in Section~\ref{sec:Salpha}.


\subsection{Populated states}

\begin{figure*}
        \centering
        \includegraphics[width=18cm]{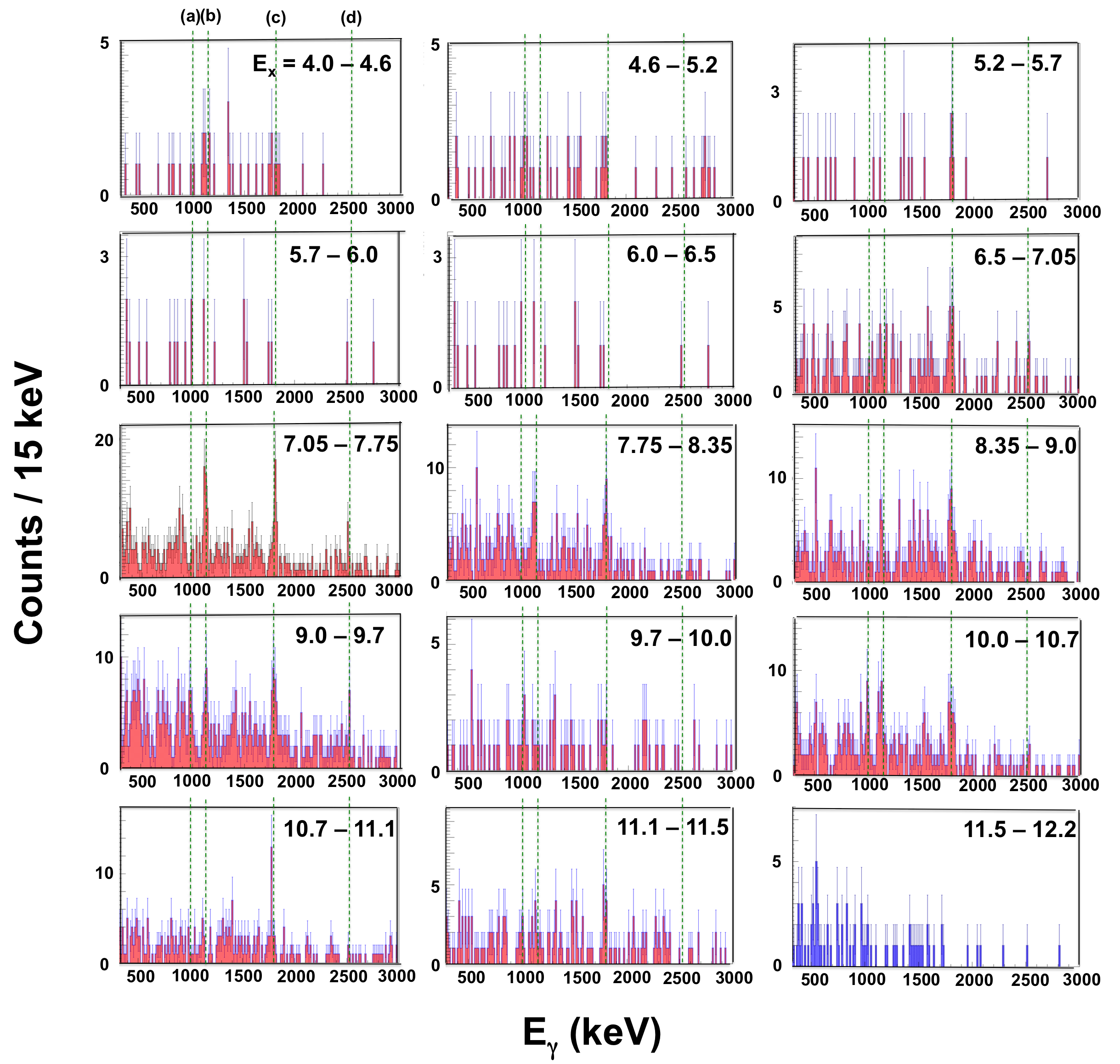}     
        \caption{Coincident $({}^{6}\mathrm{Li}, d)$ $\gamma$-ray spectra, gated on specific excitation energy ranges in $^{26}$Mg. Note the last panel (the blue-shaded histogram) was obtained by gating on $^{25}$Mg recoils instead of $^{26}$Mg. Except for the last panel, vertical dotted lines indicate the energies of major transitions from the low-lying states in $^{26}$Mg: (a) 1003, (b) 1129, (c) 1808, and (d) 2510/2523 keV, respectively (see Figure ~\ref{fig:Fig9a} (b)). 
        }\label{fig:Fig9}
\end{figure*}

Figure~\ref{fig:Fig9} shows $\gamma$-ray spectra in coincidence with specific $^{26}$Mg total excitation energy ranges. 
Simple Monte Carlo simulations of $\gamma$ spectra from the respective states were also made to estimate the expected counts of respective $\gamma$-ray transitions, with the level table \cite{NNDC2018} and our HPGe detector efficiency as inputs. 
Background counts by random coincidence were estimated from events located near the photopeaks of interest.

\subsubsection{$E_x$=4.0 -- 4.6 MeV}

Possible states making up the peak in this region are 4.318 (4$^+$) or 4.332 MeV (2$^+$). 
The possible $\gamma$ transitions from states in this region are therefore (4318$\rightarrow$1808$\rightarrow$G.S.) (99\%) and (4332$\rightarrow$1808$\rightarrow$G.S.) (79\%)
/ (4332$\rightarrow$2938$\rightarrow$1808$\rightarrow$G.S.) (15\%), where percentages in parenthesis denote the $\gamma$-decay branching ratios from the state. 
Since there seems to be the 1394 keV $\gamma$ transition from 4332$\rightarrow$2938 keV, as well as the subsequent 1129 keV transition from 2938$\rightarrow$1808 keV, it is most likely that the peak exclusively consists of the 4332 keV state. 
There are ${\sim} 150$ total deuteron events in the peak, and these two coincident $\gamma$-ray transitions both contain 3--5 counts after background subtraction. These rates are slightly higher than, but statistically consistent with, the expected 1--2 counts in these peaks after accounting for branching ratios and detection efficiencies.
From both states, 3--5 counts of another major transition (2510/2523 keV) are expected. No such transitions are, however, observed. 
The assignment of the 4.332 MeV state is supported by the angular distribution, which is consistent with a 2$^+$ DWBA calculation (see Figure~\ref{fig:Fig10}).

\subsubsection{$E_x$=4.6 -- 5.2 MeV}

There is an evident peak around $E_x$=4.901 MeV (4$^+$), which could be contaminated by $E_x$=4.835 MeV (2$^+$) or 4.972 MeV (0$^+$).
Figure~\ref{fig:Fig9} shows the coincident $\gamma$ spectrum. 
While 4.835 and 4.970 MeV states have the major 1129 keV transition (2938$\rightarrow$1808) with expected 6--7 counts, this line is not present in the $\gamma$-ray spectrum. 
Therefore, the 4.901 MeV state is likely the main contributor to the peak. This is supported by our angular distribution, which is consistent with 2$^+$ or 4$^+$ calculations. 
We note that the 1808 keV $\gamma$-transition (1808$\rightarrow$GS) is expected from any of the candidate states with $\sim$100\% probability. 
The observed number of 1808 keV events ($\sim$5 counts) is consistent with the expectations for this peak (5--6 counts for any of the states).

\subsubsection{$E_x$=5.2 -- 6.0 MeV}

There is a small peak on the left side of this region, which we assign to the $E_x$=5.476 MeV  (4$^+$) state reported by Ref.~\cite{Anantaraman1977}. No clear evidence for the population of this state can be found in the $\gamma$-ray spectrum; however, the angular distribution is consistent with the 4$^+$ calculation.

Another small peak (on the right side of this region) is assigned to the 5.715 MeV state (4$^+$), also reported by Ref.~\cite{Anantaraman1977}. For this peak, no decisive $\gamma$-ray transition was observed due to poor statistics. Our measured angular distribution agrees poorly with the 4$^+$ calculation; however, the same inconsistency is observed in Ref.~\cite{Anantaraman1977}, where it is attributed to possible multi-step reaction contributions.

\subsubsection{$E_x$=6.0 -- 7.0 MeV}

There are no states populated between $E_x$=6.0 -- 6.5 MeV, which agrees well with Ref.~\cite{Anantaraman1977}. 
%
A coincident 1808 keV $\gamma$-ray transition is clearly seen in coincidence with the excitation range of 6.5--7 MeV. This supports population of the $E_x$=6.876 MeV (3$^-$) state, which dominantly decays via the 1808 keV transition. This assignment is further supported by agreement of the angular distribution with the 3$^-$ calculations.
It is also worth mentioning that Ref.~\cite{Anantaraman1977} observed states at 6.63 and 7.06 MeV  as well, but with much weaker intensity than the 6.876 MeV. Evidence for a $6.63$ MeV shoulder is present in the deuteron spectrum, although no clear $\gamma$~rays from this state are observed.
We also note that there appears to be a $\gamma$-ray transition around 1600 keV in coincidence with this excitation region, even though no known states populated by ($^6$Li,$d$) reactions are expected to generate this $\gamma$-ray \cite{Glatz1986, NNDC2018}.  

\subsubsection{$E_x$=7.0 -- 7.7 MeV}

The large peak observed  in this region by this and other ($^6$Li, $d$) experiments  \cite{Anantaraman1977, Talwar2016} could arise from any of the 7.348, 7.371, or 7.396 keV states ~\cite{NNDC2018}.
The coincident $\gamma$-ray spectrum shows two strong transitions at 1808 and 1129 keV (i.e., decay through the $E_x$=2.938 MeV state), with another peak at 2510 keV. 
The 7.371 and 7.396 MeV states have dominant $\gamma$-ray transitions at 1003 keV and 1680 keV (expected $\sim$11 and 17 counts each), respectively. Neither of these transitions are observed, leading to the conclusion that the peak is dominated by the 7.348 MeV state (3$^-$), together with the consistency between observed and expected 1129, 1808 keV, and 2510 keV counts for this state ($\sim$30, $\sim$30, and $\sim$7 counts, respectively). However, the observed angular distribution is not consistent with the 3$^-$ calculation. Therefore, an unambiguous assignment is not possible in this region.

\subsubsection{$E_x$=7.7 -- 8.3 MeV}

There is a peak at $E_x$=8.04 MeV, which is likely a doublet consisting of the 8.034 (2$^+$) and 8.201 MeV (6$^+$) states seen by Anantaraman \emph{et al}.\ \cite{Anantaraman1977} and Talwar \emph{et al}. \cite{Talwar2016}. 
The observed angular distribution supports both 2$^+$ and 6$^+$ assignments; however, the $\gamma$-ray spectrum indicates that the 8.034 MeV state is dominant.
Two $\gamma$-ray transitions are strongly observed: 1129 and 1808 keV, and additionally there is evidence for a 511 keV transition. The presence of the 511 keV line would be expected from the high-energy (8034$\rightarrow$2938) transition from the 8.034 MeV state. At the same time, the lack of a 2510 keV transition suggests an absence of the 8.201 MeV state, which is expected to give a 2510 keV line with ${\sim} 50\%$ the intensity of the 1808 keV line. Similarly, the roughly equal 1129 and 1808 keV intensities observed in the spectrum are not consistent with decay from the 8.201 MeV state, which is expected to produce the 1129 keV line with only $16\%$ as many counts as the 1808 keV transition.

The absence of the 8.201 MeV state is inconsistent with the results of Talwar \emph{et al}.; however, this may be due to the higher beam energy used in that study, which increases population of high-spin states.

%

\subsubsection{$E_x$=8.3 -- 9.0 MeV}

There are two peaks at $E_x$=8.63 and 8.93 MeV observed by Talwar \emph{et al}. \cite{Talwar2016}, and seemingly Anantaraman \emph{et al}. \cite{Anantaraman1977} as well. 
The $\gamma$-ray spectrum shows some possible transitions, including the 1129 and 1808 keV. 
Presence of the $E_x$=8625 keV state (5$^-$) and 8931 keV state (1$^-$ or 2$^+$) both explain these transitions well. 
The observed angular distribution is inconsistent with the 8625 (5$^-$) assignment, while it is consistent with the 8931 (2$^+$) assignment. 

\subsubsection{$E_x$=9.0 -- 9.7 MeV}

In this region there is a strong peak observed by Talwar \emph{et al}.\ \cite{Talwar2016}, Giesen \emph{et al}.\ \cite{Giesen1993}, and Ugalde \emph{et al}.\ \cite{Ugalde2007}. 
Possible states in this region are 9.325 (Ugalde), 9.371, and 9.383 (Talwar) MeV. 
Looking at the $\gamma$-ray spectrum, the number of observed 1129 keV and 1808 keV counts are almost the same. Hence,
it is probable that a large fraction of $\gamma$ transitions occur via the 2938 keV state.
There is also a noticeable transition from either the 2510 or 2523 keV $\gamma$~ray, which indicates decay via the 4318 or 4331 keV state ($\rightarrow$1808 or $\rightarrow$2938$\rightarrow$1808). 

If the $({}^{6}\mathrm{Li}, d)$ peak is dominated by the 9.371 MeV state, a 1003 keV $\gamma$-ray transition (3941$\rightarrow$2938) is expected to be observed, with similar intensity to the 1129 keV and 1808 keV transitions. This is seen in the spectrum; thus it is possible the 9.371 MeV state is a significant contributor to the peak.
If the peak is dominated by the $E_x$=9.325 MeV state, the 2510/2523 keV $\gamma$-ray transition is expected with  fewer counts than a dominant $E_x$=9.383 MeV state (expected ${\sim}5$ and ${\sim}$10 counts, respectively). However, due to the low numbers of expected and observed counts, we cannot make a definitive distinction between these two states on this basis.

Turning to the angular distributions, the observed shape 
supports a 2$^+$ 
assignment. This tentatively favors a dominant 9.325 MeV state, which is reported to have spins in the ranging from 2$^+$ to 4$^+$ \cite{NNDC2018}. Conversely, the 9.383 MeV state claimed by Talwar \emph{et al}. was reported having $J^{\pi}$$=$0$^+$ or 1$^-$ \cite{Talwar2016},  or 6$^+$ \cite{NNDC2018}, and the 9.371 MeV state was reported having 4$^+$ \cite{NNDC2018}, neither of which (although 0$^+$ and 1$^-$ are not shown in the figure) is not consistent with the present data.

\subsubsection{$E_x$=9.7 -- 10.0 MeV}

While Talwar \emph{et al}. and Giesen \emph{et al.} both report a state at $E_x$=9.99 MeV, we instead observed a peak at 9.856 MeV. 
The angular distribution supports a 2$^+$ assignment, which agrees with our assignment to the 2$^+$, 9.856 MeV state \cite{NNDC2018}. No clear $\gamma$-ray transitions were observed in this region.

\subsubsection{$E_x$=10 -- 10.7 MeV}

In this excitation region, clear $\gamma$-ray peaks are observed at 1003, 1129, and 1808 keV. 
The states observed in past ($^6$Li, $d$) measurements are 10.36, 10.57, 10.70 MeV \cite{Giesen1993, Talwar2016} (note that Talwar \emph{et al}. \cite{Talwar2016} did not observe the 10.57 MeV state). 
For the 10.57 MeV state, reliable $\gamma$-ray transition data are also available from a ($\gamma, \gamma^\prime$) experiment \cite{Longland2012}. 
Possible candidate states around the 10.36 MeV region are 10.362 and 10.377 MeV \cite{NNDC2018}.
The 10.362 MeV state decays dominantly via the 10362$\rightarrow$1808$\rightarrow$GS scheme with some weaker schemes, generating a $\sim$100\% 1808 keV $\gamma$~ray which is several times more intense than the next strongest transition (1129 keV).
The 10.377 MeV state can emit a 1003 keV $\gamma$-ray with some strength ($\sim$25\%) after transitioning through the 5.715 MeV state (5.715$\rightarrow$4.350$\rightarrow$1.808$\rightarrow$GS or 5.715$\rightarrow$3.941$\rightarrow$2.938$\rightarrow$1.808$\rightarrow$GS). Hence both are reasonably consistent with the observed $\gamma$-ray spectrum.

The 10.57 MeV state reported by Longland \emph{et al}. \cite{Longland2012} decays through the 4.972 MeV state, generating 1129, 1808, 2003 keV $\gamma$ rays. For this decay scheme, the   2003 keV transition is expected to have nearly half the intensity of the 1808 keV transition. We do not observe a 2003 keV transition at all, making it likely that the  
10.57 MeV state is only weakly populated. This agrees with both Giesen \emph{et al}.\ \cite{Giesen1993} and Talwar \emph{et al}.\ \cite{Talwar2016}.
 
The counts around 10.70 MeV can arise from two possible candidate states.
The first is at 10.707 MeV and decays via the 5.715 MeV state. The second is at 10.693 MeV that also decays through the 5.715 (10.693$\rightarrow$7.396$\rightarrow$5.715). 
Thus, the two states generate similar $\gamma$-ray spectra, except for the 1680 keV transition from (7.396$\rightarrow$5.715) in the 10.693 MeV state.  
Both states generate a 1129 and 1003 keV transition (via the 2.938 and 3.941 MeV states, respectively), so the observed $\gamma$-ray transitions could be attributed to either candidate.
However, since the 1680 keV transition is not observed, the 10.707 MeV state is more likely as a candidate for the 10.70 MeV state populated by other ($^6$Li,$d$) experiments. 

Taking the observed $\gamma$-rays into consideration, it is likely that multiplets near both 10.36 MeV and 10.7 MeV are present. 
However, the identifying $\gamma$-ray transitions are not seen, e.g., the 1003 keV transition, in coincidence with the $E_x=10.7$--11.1 MeV region (where ${\sim}$half of the strength of a state around 10.7 MeV is expected to lie). This indicates that population of states near 10.7 MeV is weak.
%
This is supported by both Giesen \emph{et al}.\ and Talwar \emph{et al}., who report the 10.36 MeV as being the strongest of the three observed states. 

Further assignment to specific states is not possible given the energy resolution and ambiguous origin of the observed $\gamma$ rays. Similarly, we made no attempt to extract angular distributions due to the complicated and uncertain mix of multiple states in this region.

\subsubsection{$E_x$=10.7 -- 11.1 MeV}

States observed in the past ($^6$Li, $d$) experiments are the 10.805 (or 10.823) and 10.949 MeV. 
The $\gamma$-ray spectrum shows only a single strong transition at 1808 keV.
This indicates there there is a direct transition from the initial state to the 1808 keV state. 
This agrees with both the 10.805 and 10.949 MeV states observed by Longland \emph{et al}.\ \cite{Longland2012} (78\% and 95\% via the 1808 keV, respectively). It also rules out the possibility of the 10.823 MeV state being strongly populated since comparable 1129 and 1808 keV transitions are expected from that state, 
according to the findings of Lotay \emph{et al}.\ \cite{Lotay2019}, who observed that the 10.823 MeV state decays via the 5.476 MeV state, which should generate 1157, 1808, and 2510 keV $\gamma$ rays (assuming the 10.823$\rightarrow$5.746 MeV transition is dominant). In contrast, only the 1808 keV line is present in our spectrum.

It was not possible to draw any further conclusions about the relative intensities of the 10.805 and 10.953 MeV states composing the peak. Similarly, we did not attempt to construct an angular distribution due to the uncertain mix of states in this region.


\subsubsection{$E_x$=11.1 -- 11.5 MeV}

This region includes excitation energies above the neutron decay threshold, and here  (in the lower-middle panel of Fig.~\ref{fig:Fig9}) we show only $\gamma$-ray transitions in coincidence with ${}^{26}$Mg recoils. 
The astrophysically-important peak at $E_x$=11.32 MeV in the deuteron spectrum can be clearly observed, consistent with Giesen \emph{et al}.\ and Talwar \emph{et al}. 
According to past measurements ~\cite{Wolke1989,NNDC2018}, this state decays directly via the 1808 keV state (11.318$\rightarrow$1.808) with 47$\pm$4\%, through the 11.318$\rightarrow$7.060$\rightarrow$GS scheme (19$\pm$1\%), or through the 11.318$\rightarrow$7.060$\rightarrow$3588 scheme (18$\pm$1\%), or 11.318$\rightarrow$7.060$\rightarrow$2938 scheme (5.3$\pm$0.4\%), or 11.318$\rightarrow$7.060$\rightarrow$1808 scheme (11$\pm$0.8\%).
All together, the expected 1808 keV $\gamma$-ray probability from this state is 81$\pm$5\%. Considering the $\gamma$-ray branchings, the ${\sim}170$ deuteron events in this peak in coincidence with ${}^{26}$Mg recoils, and the 5$\%$ $\gamma$-ray detection efficiency at 1808 keV, we expect to see 7$\pm$1 1808 keV counts from the 11.32 MeV state. This is consistent with the  10$\pm$3 counts seen for this peak. The next-most-probable transition from the 11.32 MeV state is the 1129 keV line (2938$\rightarrow$1808). This has an intensity $<$10\% of the total 1808 keV intensity and hence should not be observable above background, which is consistent with our spectrum ($<$ 3 counts after background subtraction). 
%

Concerning other candidate peaks in this region, Lotay \emph{et al}.\ observed that the $E_x$=11.17 MeV level decays via the 5.476 MeV state, which eventually leads to 1129 (69\%), 1157 (51\%), 2510 (51\%), and 1808 keV ($\sim$100\%) $\gamma$~rays \cite{Lotay2019}. Since we do not observe measurable 1129, 1157 keV and 2510 keV transitions (2--3, 2--3 and 0 counts at most, respectively), it appears that this state is weakly populated 
compared to the 11.32 MeV. This conclusion is consistent with our earlier publication \cite{Ota2019}, which set an upper limit of 0.15 on the $\alpha$ spectroscopic factor from an analysis of the ${}^{25}$Mg-gated deuteron spectrum (without considering $\gamma$~rays).
However, it should be noted that the branching ratio of the transition (11.17 MeV$\rightarrow$5.476 MeV) is unknown. 
Thus, assuming the possibility that the decay mode is not dominant, it is not possible to draw a firm conclusion. 
Lotay \emph{et al}.\  also reported that the $E_x$=11.08 MeV state decays via the 4.318 MeV state, which eventually generates 2510 and 1808 keV $\gamma$~rays. 
We do not observe a transition at 2510 keV. Although branching ratios of the respective decay transitions are not available, assuming the decay via the 4.318 MeV state is dominant,  this state appears to also be only weakly populated compared to the $E_x$=11.32 MeV. 
This is consistent with both Giesen \emph{et al}. and Talwar \emph{et al}., who also failed to observe this state \cite{Giesen1993, Talwar2016}. However, the state was strongly populated and observed in the sub-Coulomb ($^6$Li, $d$) and ($^7$Li, $t$) measurements of Jayatissa \emph{et al} \cite{Jayatissa2019}.

\subsubsection{$E_x$=11.5 -- 12.2 MeV}
\label{ss:Mg25CoincGamma}

Peaks in this region are exclusively neutron unbound and hence the corresponding $\gamma$-ray spectrum (lower-right panel of Fig.~\ref{fig:Fig9}) is gated on ${}^{25}$Mg-$d$ coincidences.
In particular, we note that the strongly-populated peak at $E_x = 11.83$ MeV comes in coincidence with a 585 keV $\gamma$-ray transition from the first excited state of ${}^{25}$Mg by ($^6$Li,$d n_1\gamma$). 
We also note that the 585 keV  state is an isomer with a $3.38$ ns half life \cite{NNDC2018}. 
Therefore, the average position of $^{25}$Mg ions when the $\gamma$-rays are emitted is $\sim$10 cm off from the target position along the beam direction ($z$). 
Given the dimension of our HPGe clover detectors ($z=\pm5$ cm), a large portion of the $\gamma$-rays are emitted outside the detector array. 
As a result, the efficiency for observing the delayed $\gamma$~rays is significantly lower than for prompt decays. 
Nevertheless, $\sim$5 counts were observed for the transition, which is significantly more than the expected ${\sim}$0.4 counts based on observed $\sim$550 counts of deuterons, the reported $n_1$/$n_0$ ratio of $\sim$0.01 ~\cite{Wolke1989}, and 8\% $\gamma$-ray efficiency for prompt $\gamma$-rays (note we have not attempted to calculate the reduction in efficiency resulting from the $3.38$ ns half life). 
This suggests that either the  $n_1$/$n_0$ ratio for this state is significantly larger than reported in Ref.~\cite{Wolke1989}, or that there are multiple states besides the 11.83 MeV state populated in this region.

%


\subsection{Relative $\alpha$ strength}
\label{sec:Salpha}

As mentioned, the $\alpha$ partial decay width, $\Gamma_\alpha$, of $\alpha$-unbound resonances within the Gamow window is the most important  parameter determining the total ${}^{22}\mathrm{Ne} + \alpha$ stellar reaction rate. For a given spin and resonance energy, this parameter is proportional to the $\alpha$ spectroscopic factor,  $S_\alpha$. The spectroscopic factor is also an indicator of possible $\alpha$ cluster structures in both bound and unbound states in ${}^{26}$Mg. For a given spin, the spectroscopic factor can be extracted from the angular distributions shown in Fig.~\ref{fig:Fig10}.
Table~\ref{tab:tab4} shows the extracted spectroscopic factors of the states which were clearly observed in the present experiment. 
%
Between excitation energies of 5.7--9.3~MeV, this is the first time that individual states have been identified and assigned spectroscopic factors.

As in Ref.~\cite{Ota2019}, the $S_\alpha$ presented in Table~\ref{tab:tab4} are normalized to the spectroscopic factor for the state at $E_x = 11.318$~MeV. The $S_\alpha$ for this state was calculated from the ratio $\Gamma_\alpha / \Gamma_\alpha^{\mathrm{(s.p.)}}$, where the single-particle $\alpha$ width was calculated numerically from the wave function used in the DWBA calculation. The $\alpha$ partial width was taken from the results of direct $(\alpha,\gamma)$ measurements, $\Gamma_\alpha =  \omega \gamma_{(\alpha,\gamma)} (1 + \Gamma_n/\Gamma_\gamma) / (2J+1)$, with $\omega \gamma_{(\alpha,\gamma)} = 37 \pm 4~\mu$eV (the weighted average of Refs.~\cite{Wolke1989, Hunt2019})
 and $\Gamma_n/\Gamma_\gamma  =  1.14 \pm 0.26$ from Ref.~\cite{Ota2019}. Separate normalizations assuming $J^\pi_{11318} = 0^+$ and $J^\pi_{11318} = 1^-$ were calculated and are presented in Table~\ref{tab:tab4}.
A separate normalization was also performed to the excited state at 4.332~MeV, which was reported as having $S_\alpha / S_\alpha^\mathrm{(g.s.)} = 0.29$ in Ref.~\cite{Anantaraman1977}. For this normalization, $S_\alpha^\mathrm{(4332)}$ was fixed at 0.29, effectively giving the same normalization as Ref.~\cite{Anantaraman1977}, which fixed $S_\alpha^\mathrm{(g.s.)} = 1.0$. 


The spectroscopic factors and associated $\alpha$ partial widths (or upper limits) for states from 11.11--11.32~MeV have already been discussed in Ref.~\cite{Ota2019}. 
Between 5.7--9.3~MeV, the presently determined spectroscopic factors are reported for the first time. We note the presence of a number of possible strong $\alpha$ cluster states ($S_\alpha \sim 0.1$--0.5) in this region. More detailed conclusions about these states would require additional studies that focus more directly on this region.
For excitation energies below $5.7$ MeV, we can make direct comparisons to Ref.~\cite{Anantaraman1977} when normalizing to $S_\alpha^{\mathrm{(g.s.)}} \equiv 1$. In this region, the spectroscopic factor for the $E_x$=5.475 MeV state agrees with Ref.~\cite{Anantaraman1977},  while spectroscopic factors for the $E_x=4.835$ and $4.901$~MeV states are 2--3 times  larger.
This latter difference likely results from the present treatment of the observed peak as a single state. In contrast,  Ref.~\cite{Anantaraman1977} treated their peak as a triplet and obtained separate $S_\alpha$ for individual states using a least-squares fit to their angular distribution. We note that the ratios of differential cross sections to the 4.3 and 4.9 MeV states at $\theta_{CM} \sim 10^\circ$, are consistent for the two experiments. As a result, we can claim reasonable agreement between observed spectra for states in this region, although the interpretations differ.


Above $E_x = 11.32$~MeV, we focus on the $\alpha$ partial width extracted for the strongly-populated resonance at $E_x = 11.83$~MeV. This resonance is the dominant contributor to the ${}^{22}\mathrm{Ne}(\alpha,n){}^{25}\mathrm{Mg}$ rate in the high-temperatures ($T \sim 1$ GK) realized during the C-shell burning phase of the $s$-process. This state has been observed in a number of past direct $(\alpha,n)$ measurements,
which report values of  $610(90)$ meV \cite{Wolke1989}, $1067(42)$ meV \cite{Jaeger2001}, and $1105(120)$ meV \cite{Drotleff1993}.
These results are in poor statistical agreement ($\chi^2/\mathrm{NDF} = 11$). Hence, we calculate an inflated weighted average of the three measurements,  $930(170)$~meV, as prescribed by Longland \emph{et al}. \cite{Longland2012} and use this value for comparisons with the present data.
%

In Table~\ref{tab:tab5}, we show the resonance strength of this state extracted from the present data, calculated as $\omega\gamma \simeq  (2J+1)\Gamma_\alpha$. The calculations were done assuming both $0^+$ and $1^-$ spin-parity for the $E_x=11.32$ MeV state (used for normalization) and for separate $J^\pi$ values of $J^\pi = (1^-,2^+,3^-)$ as reported in Ref.~\cite{Wolke1989}.
Taken at face value, the resonance strengths support a $2^+$ assignment to the 11.83 MeV state when compared with the direct-measurement strength of 930(170) meV.  
We stress, however, that these calculations assume only the 11.83 MeV state exists in the observed $({}^{6}\mathrm{Li}, d)$ peak, whereas both the width of the observed peak and the intensity of coincident 585 keV $\gamma$~rays (Section~\ref{ss:Mg25CoincGamma}) indicate other states being populated in this energy region---for example at $E_x$=11.89 or 11.93 MeV. A similar conclusion was also made about the spectrum observed in Ref.~\cite{Giesen1993}. As a result, the resonance strengths reported in Table~\ref{tab:tab5} are most conservatively treated as upper limits. Taken as such, comparison with the direct measurements rules out the $3^-$ assignment and leaves a possibility for $1^-$ or $2^+$.  The $1^-$ possibility is also consistent with the presently-observed angular distribution (see Fig.~\ref{fig:Fig10}).

\begin{table*}
\caption{\label{tab:tab4} Excitation energies, spin-parities, and spectroscopic factors for 
$^{26}$Mg states populated in the present ($^6$Li,$d$) experiment. Three separate normalizations are used for the spectroscopic factors as explained in the text. When available, spectroscopic factors from Refs.~\cite{Giesen1993, Talwar2016, Anantaraman1977} are also listed.
}
\begin{ruledtabular}
\begin{tabular}{cccccccc}
$E_x$ (keV) & $J^\pi$ & $S_\alpha (J_{11318}=0)$ & $S_\alpha (J_{11318}=1)$ & $S_\alpha/S_{\alpha}^{,\mathrm{(g.s.)}}$ & Refs.~\cite{Giesen1993, Talwar2016, Anantaraman1977} \\
\hline
$4332{\footnotemark[1]} $  
         & $2^+${\footnotemark[6]} & $0.08(1)$ & $0.14(2)$ & 0.29(4){\footnotemark[11]} & 0.29(4){\footnotemark[12]}  \\
$4835,4901{\footnotemark[1]} $  
         & $2^+,4^+${\footnotemark[6]} & $0.060(10), 0.12(1)$ & $0.10(1), 0.21(2)$  & 0.22(2), 0.45(4) & 0.06(2){\footnotemark[12]}, 0.20(4){\footnotemark[12]} \\
$5476{\footnotemark[1]}$  
         & $4^+${\footnotemark[6]} & $0.031(5)$ & $0.054(10)$& 0.11(2) & 0.08{\footnotemark[12]}  \\
$5715{\footnotemark[1]}$  
         & $4^+${\footnotemark[6]} & $0.020(3)$ & $0.034(3)$& 0.07(1)  & \\
$6876{\footnotemark[1]}$  
         & $3^-${\footnotemark[6]} & $0.11(1)$ & $0.19(2)$& 0.40(4) &   \\
$7365, 7396{\footnotemark[1]}{\footnotemark[3]}$  
         & $2^+${\footnotemark[7]}, $(5^-)${\footnotemark[6]} & $0.30(3), 0.30(3)$ & $0.52(5), 0.52(5)$ & 1.12(12), 1.12(12) & \\
$8036{\footnotemark[1]}{\footnotemark[3]}$  
         & $2^+${\footnotemark[7]} & $0.12(1)$ & $0.21(2)$ & 0.45(4) &\\
$8201{\footnotemark[1]}{\footnotemark[3]}$  
         & ($6^+)${\footnotemark[6]} & $0.17(2)$ & $0.29(3)$ & 0.62(8) & \\         
$8625{\footnotemark[1]}{\footnotemark[3]}$  
         & $5^-${\footnotemark[6]} & $0.10(1)$ & $0.17(2)$ & 0.36(4)  &\\
$8931{\footnotemark[1]}{\footnotemark[3]}$  
         & $2^+${\footnotemark[7]} & $0.08(1)$ & $0.15(2)$ & 0.31(4) & \\
$9325, 9371, 9383{\footnotemark[1]}{\footnotemark[2]}{\footnotemark[3]}{\footnotemark[4]}$  
         & $2^+$ & $0.21(2)$ & $0.38(4)$ & 0.80(8)  &\\
$9856$
         & $2^+${\footnotemark[6]} & $0.06(1)$ & $0.10(1)$ & 0.22(4) &\\
$10805+10949{\footnotemark[2]}{\footnotemark[3]}{\footnotemark[4]}{\footnotemark[5]}$  
         & $1^-${\footnotemark[8]} & $0.12(1)$ & $0.21(2)$ & 0.45(4) & $< (0.06+0.15)${\footnotemark[13]} \\                   
$11318{\footnotemark[2]}{\footnotemark[3]}{\footnotemark[5]}$  
         & $0^+,1^-${\footnotemark[9]} & $0.31(5)$ & $0.18(3)$ & 1.16(12), 0.67(8) & 0.04{\footnotemark[14]}, $0.43${\footnotemark[13]}  \\                                    
$11831{\footnotemark[2]}$  
         & $1^-,2^+${\footnotemark[10]} & $0.27(3), 0.18(2)$ & 0.45(5), 0.31(3) & 1.00(12), 0.67(8) & 0.20{\footnotemark[14]}, 0.11{\footnotemark[14]}  \\
\end{tabular}
\end{ruledtabular}
\footnotetext[1]{Observed by \cite{Anantaraman1977}. $E_x$$>$5.715 MeV are observed in this work by digitizing the spectrum from \cite{Anantaraman1977}. }
\footnotetext[2]{Observed by \cite{Giesen1993}}
\footnotetext[3]{Observed by \cite{Talwar2016}}
\footnotetext[4]{Observed by \cite{Ugalde2007}}
\footnotetext[5]{Observed by \cite{Jayatissa2019}}
\footnotetext[6]{Excitation energies and spin-parities are from \cite{NNDC2018}}
\footnotetext[7]{Excitation energies and spin-parities are from \cite{Talwar2016}}
\footnotetext[8]{Excitation energies and spin-parities are from \cite{Longland2012}}
\footnotetext[9]{Excitation energies and spin-parities are from \cite{Ota2019}}
\footnotetext[10]{Excitation energy is from \cite{Giesen1993}}
\footnotetext[11]{normalized to \cite{Anantaraman1977}, where $S_\alpha^{\mathrm{(g.s.)}} = 1$. }
\footnotetext[12]{from \cite{Anantaraman1977}, normalized to $S_\alpha^{\mathrm{(g.s.)}} = 1$. }
\footnotetext[13]{from \cite{Talwar2016}}
\footnotetext[14]{from \cite{Giesen1993}}
\end{table*}

\begin{table}
\caption{\label{tab:tab5} ${}^{22}\mathrm{Ne}(\alpha, n){}^{25}\mathrm{Mg}$ resonance strengths for the $E_x$=11.83 MeV state in ${}^{26}$Mg, deduced from the present data.  Resonance strengths are presented in units of meV.
The subscripts to the $\omega\gamma$ symbol deonte the assumed $J^\pi$ assignment for the 11.83 MeV state. Calculations for both $0^+$ and $1^-$ spin-parities of the $E_x = 11.32$ MeV state are included as indicated. For comparison, the inflated weighted average of direct-measurement resonance strengths  $\omega\gamma = 930(170)$ meV.
}
\begin{ruledtabular}
\begin{tabular}{cccc}
 $J^\pi_{1132}$ & $\omega\gamma_{1^-}$  & $\omega\gamma_{2^+}$  & $\omega\gamma_{3^-}$  \\
\hline
$0^+$  
         & $2634(606)$ & 730(168) & 194(45)\\
$1^-$  
         & $4380(1007)$ & 1215(279) & 323(74)\\
\end{tabular}
\end{ruledtabular}
\end{table}



\section{Astrophysical implications}

To investigate the impact of our recent measurements on $s$-process nucleosynthesis, we have performed abundance calculations using the  
$^{22}$Ne($\alpha$,$n$)$^{25}$Mg and $^{22}$Ne($\alpha$,$\gamma$)$^{26}$Mg reaction rates presented in Ref.~\cite{Ota2019}. The calculations were performed using the post-processing nucleosynthesis code MPPNP \cite{Herwig2018}, which was  developed by the NuGrid collaboration. 
The code takes as an input the stellar evolution trajectories calculated by the one-dimensional stellar evolution code MESA \cite{Paxton2010}.
The MESA trajectories employed in the present work represent a variety of initial stellar conditions and were prepared by the NuGrid collaboration in previous works \cite{Pignatari2016, Ritter2018, Battino2019}. 
In the present calculations, we only varied the $^{22}$Ne($\alpha$,$n$)$^{25}$Mg and $^{22}$Ne($\alpha$,$\gamma$)$^{26}$Mg reaction rates and kept the values of all other input parameters fixed. As a result, the present calculations only probe the impact of the ${}^{22}\mathrm{Ne} + \alpha$ reactions on the final $s$-process abundances.

For the AGB stars with initial mass M=3 and 5$M_\odot$, we consider 
abundances 
at the stellar surface after the last Third Dredge Up episode. 
For the massive star models, we focus on the pre-supernova abundances in the middle of the convective C shell once the $s$-process nucleosynthesis has finished and before the final core-collapsed supernova (CCSN) explosion.
However, we also consider abundances at the end of the He core burning, to better understand the total nucleosynthesis.

In the first part of our calculations (Section~\ref{ss:AstroA}), 
we show the sensitivity of the predicted $s$-process to variance in the ${}^{22}\mathrm{Ne} + \alpha$ reaction rates as given in prior studies: Massimi \emph{et al} \cite{Massimi2017} (where only an upper limit was given),  Longland \emph{et al} \cite{Longland2012}, Talwar \emph{et al} \cite{Talwar2016}, and Adsley \emph{et al}.\ \cite{Adsley2021}, 
as well as our previous publication \cite{Ota2019}). 
These rates are hereafter referred to as MA17, LO12, TA16, AD21, and OT20, respectively. 
These calculations are made for 3, 5 and 25 $M_\odot$ stars with metallicity $Z = 0.02$ from the beginning of H-burning until the end of hydrostatic stellar evolution. 
The 3 $M_\odot$ model is from Ref.~\cite{Battino2019}, and the 5 and 25 $M_\odot$ models are from Ref.~\cite{Ritter2018}. 

In the second part of the calculations (Section~\ref{ss:AstroB}), we investigate the impact of hypothetical changes of selected ${}^{22}\mathrm{Ne} + \alpha$ resonances on predicted final s-process abundances.
 In particular, we show the sensitivity of the $s$-process abundances to uncertainties in the $\alpha$ strength of  three resonances ($E_x$=11.112, 11.171, and 11.319 MeV). These resonances were chosen for study because they have the potential to dominate one or both of the ${}^{22}\mathrm{Ne} + \alpha$ rates in certain temperature regimes and furthermore, because the present literature is either lacking key information or contains discrepancies between studies.
 


\subsection{Impact of new reaction rates}
\label{ss:AstroA}

\begin{figure*}
        \centering
          \includegraphics[width=17cm]{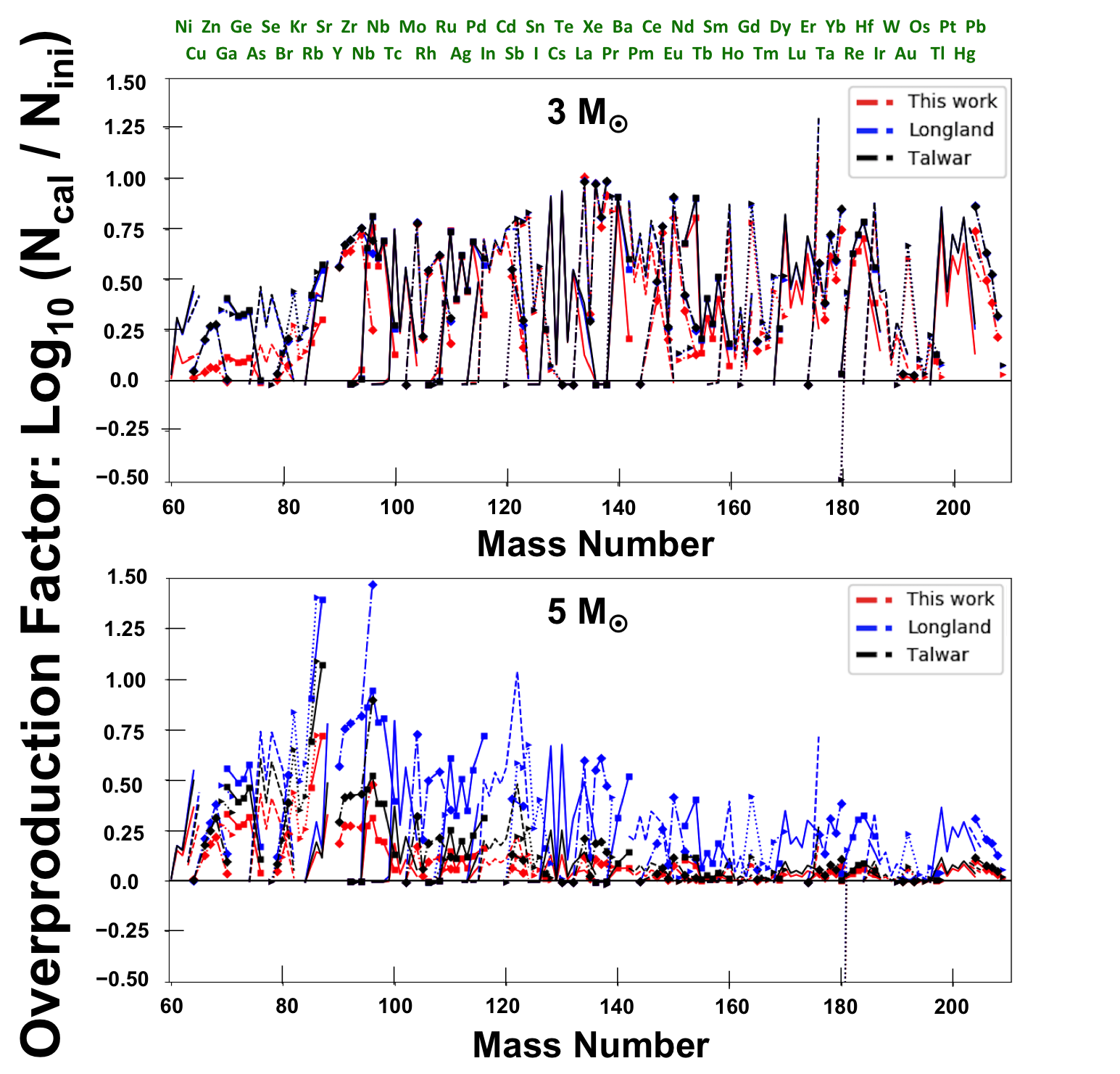}                   
        \caption{Calculated s-process overproduction factors using various available $^{22}$Ne($\alpha$,n) and $^{22}$Ne($\alpha$,$\gamma$) rates. See text for details. Isotopes of the same elements are connected by adjoining lines. 
        }\label{fig:Fig11}
\end{figure*}
\begin{figure*}
        \centering
          \includegraphics[width=17cm]{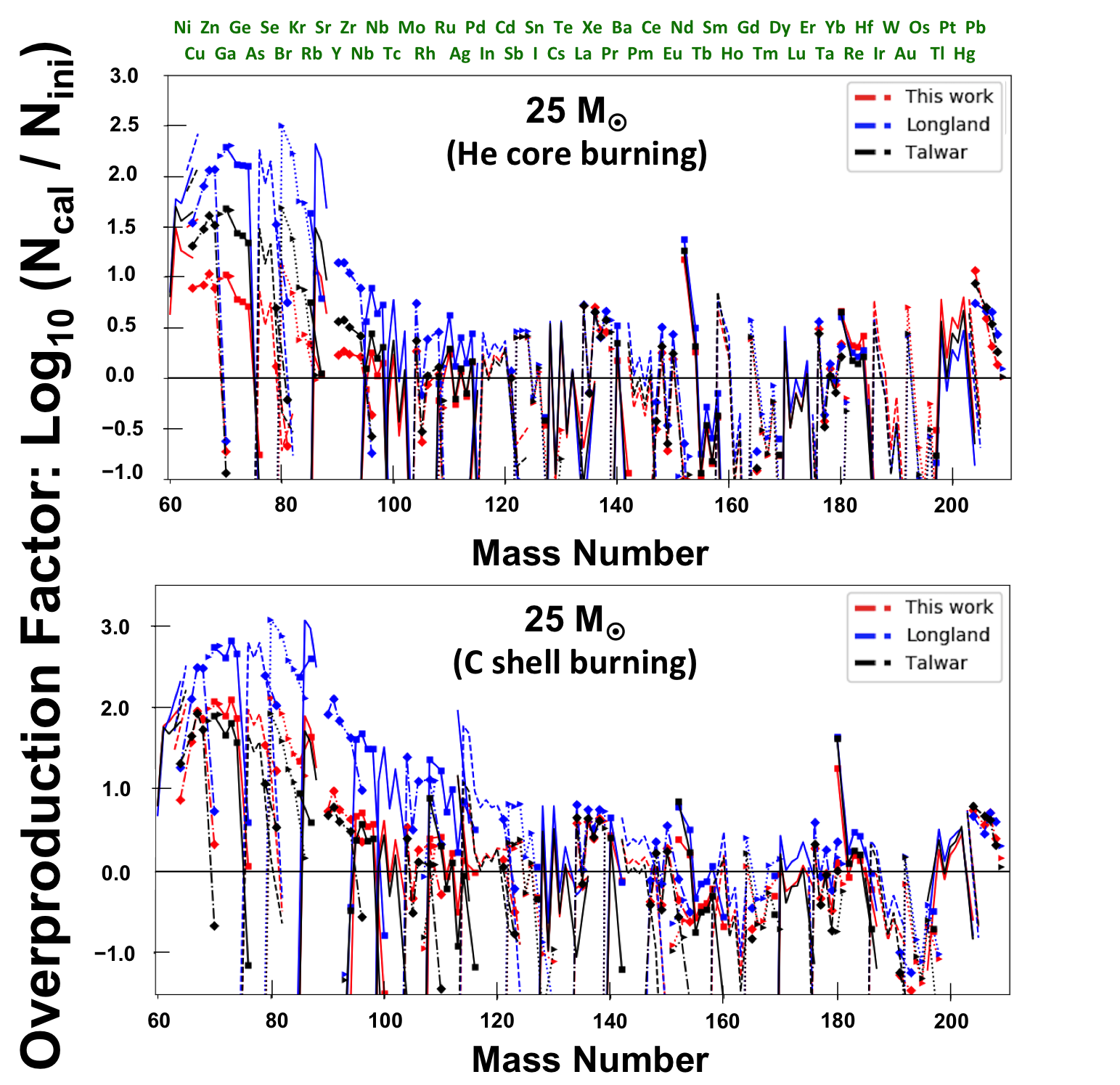}                   
        \caption{Calculated s-process overproduction factors using various available $^{22}$Ne($\alpha$,n) and $^{22}$Ne($\alpha$,$\gamma$) rates. See text for details. Isotopes of the same elements are connected by adjoining lines. 
        }\label{fig:Fig11b}
\end{figure*}

Figure~\ref{fig:Fig11} shows the isotopic abundances produced using different $^{22}$Ne+$\alpha$ reaction rates by LO12, TA16 and OT20 
for 3 and 5 $M_\odot$ with Z=0.02, respectively. For all three cases, the ``recommended'' rates published in the respective papers were used in the calculations.
Figure~\ref{fig:Fig11b} shows the same abundances as Figure~\ref{fig:Fig11} but for 25 $M_\odot$ at the end of He core burning and in the middle of C shell burning, respectively.
Abundances are shown as overproduction factors, which are defined by log$_{10}$($X_{cal}/X_{ini}$), where $X_{cal}$ and $X_{ini}$ represent the final and initial mass fractions, respectively.
The mass fraction, $X$, is given by $A\times N$ where $A$ is mass of isotope and $N$ is its abundance. The sum of all mass fractions from hydrogen up to bismuth is equal to unity. The initial abundance is solar-scaled as in Ref.~\cite{Pignatari2016}, based on Grevesse and Noels \cite{Grevesse1993} and with the isotopic ratios from Lodders \cite{Lodders2003}.

The largest 
impact from using different 
rates is generated in mass $A=60$--90 ($Z=27$--40) in all the models considered. 
The large overproduction factors of heavier elements in the 3$M_\odot$ model is predominantly driven by neutrons from $^{13}$C($\alpha,n$)$^{16}$O, not from the $^{22}$Ne($\alpha,n$)$^{25}$Mg 
reaction. 
Thus, only marginal differences are generated by using different reaction rates in the mass region ($A>$90). 
In this case, the contribution to heavy elements from $^{22}$Ne($\alpha,n$)$^{25}$Mg reaction is generally more limited to 
isotopes near s-process branching points, that are affected by its short but high neutron flux. For a comprehensive list of branching points we refer to Ref.~\cite{bisterzo:15}.

As a general trend, the LO12 rates produce the highest s-process abundances, while our rates show the lowest s-process efficiency. Using the TA16 rates, we obtain s-process yields that are somewhere in between. 
The main difference of the TA16 rate from LO12 is the treatment of $E_x$=11.17 MeV resonance.  
While LO12 did not include the resonance, TA16 assigned a large $(\alpha,\gamma)$ strength in addition to the $(\alpha,n)$ upper limit imposed by Ref.~\cite{Jaeger2001}. 
Thus, TA16 rates, especially the $(\alpha,\gamma)$ rate, are much larger than LO12 at the $s$-process temperature range. 
In the 3 $M_\odot$ star, the differences between LO12 and TA16 are minor; however, 
the 
impact becomes clearly 
visible in the 5 and 25 $M_\odot$ stars. 
This is 
because the TA16's large $(\alpha,\gamma)$ rate 
is more efficient in depleting $^{22}$Ne in competition with the $(\alpha,n)$ reaction, leading to the reduced neutron flux. 
The main difference of our rates from LO12 is the 
reduced $(\alpha,n)$ strength in $E_x$=11.32 MeV resonance, by about a factor of 3. 
This generates the lower efficiency in producing neutrons for the s-process shown in Figure~\ref{fig:Fig11} and ~\ref{fig:Fig11b} by using the $^{22}$Ne+$\alpha$ rates presented in this work. 
In the 25 $M_\odot$ star, it should be noted that the overproduction factors by TA16 nearly remain unchanged in the C shell burning compared to the He core burning, 
while the overproduction factors by LO12 and OT20 are largely enhanced. 
This is because $^{22}$Ne are drastically consumed during the He core burning 
using TA16, thus s-process in the C shell burning is largely suppressed. 

\begin{figure*}
        \centering
          \includegraphics[width=13cm]{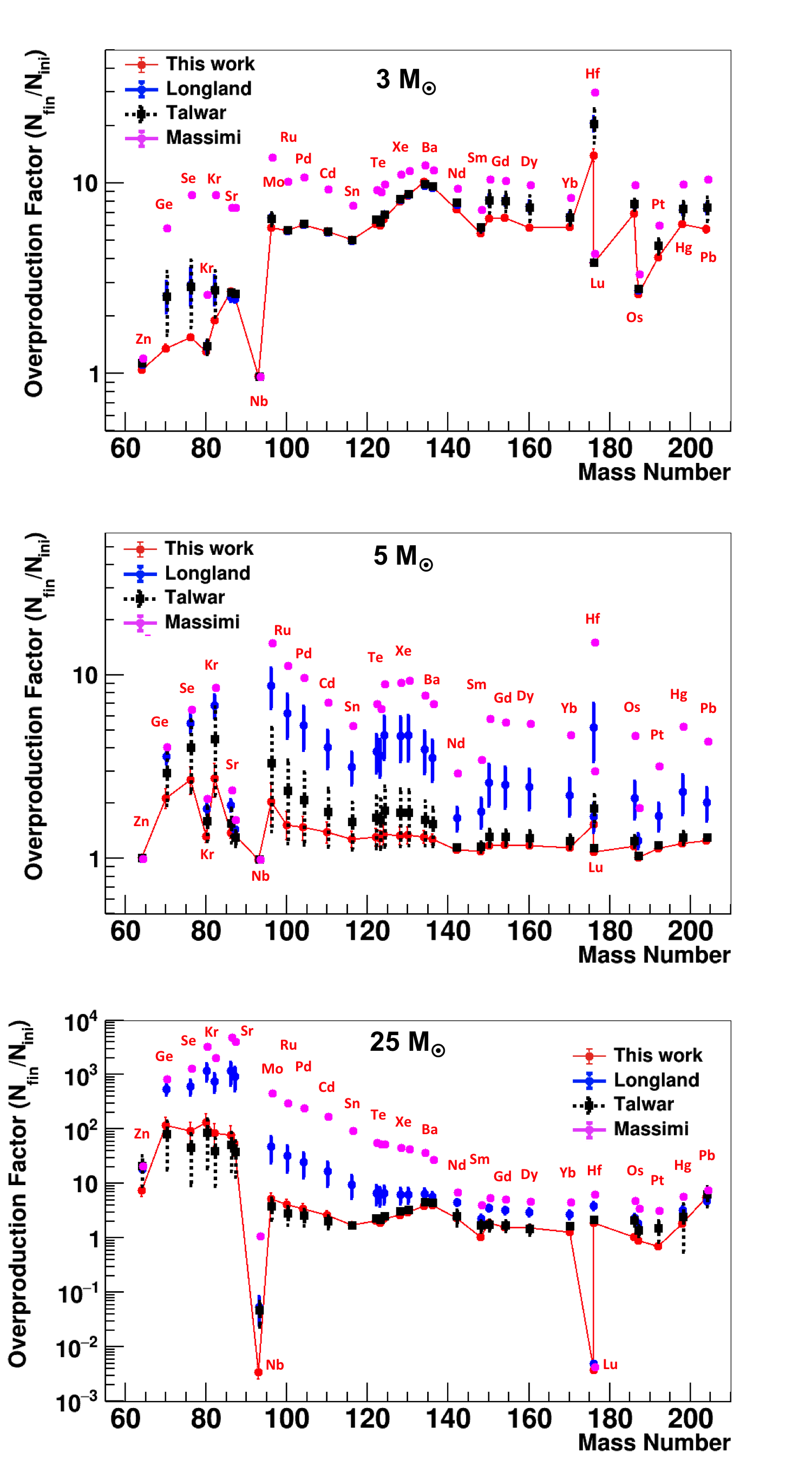}            
        \caption{Calculated s-process overproduction factors using various available $^{22}$Ne($\alpha$,n) and $^{22}$Ne($\alpha$,$\gamma$) rates, for s-only nuclei. Top, middle, and bottom panels correspond to 3, 5, and 25 (in the middle of the C shell burning) $M_\odot$ cases, respectively. Note that the rate given by Massimi \emph{et al}.\ \cite{Massimi2017} is their upper limit (see texts for details). 
        }\label{fig:Fig12}
\end{figure*}

Figure~\ref{fig:Fig12} shows the isotopic abundances of representative s-only nuclei (see e.g., \cite{Prantzos2020}), produced by using the same three rates considered for Figure~\ref{fig:Fig11} and ~\ref{fig:Fig11b}. Additionally, we have considered the upper-limit $^{22}$Ne+$\alpha$ rates given in MA17. Calculations using the AD21 rates were also performed with results indistinguishable from the TAMU rates. This is expected since both studies use very similar $^{22}$Ne+$\alpha$ rates, with only minor rate differences resulting from the treatment of low energy resonances.  As a result, the AD21 calculations are not shown in Figure~\ref{fig:Fig12}.
For all the rates excluding MA17, error bars are provided for the s-process abundances. 
Uncertainties 
are estimated by using the combination of the upper limit of the ($\alpha,n$) and of the lower limit of the $(\alpha,\gamma)$ (yielding the highest s-process efficiency), and the lower limit of the ($\alpha,n$) with the upper limit of the $(\alpha,\gamma)$ (yielding the lowest s-process efficiency). 
 
Overall, the upper limit by MA17 is consistent with all the other rates.  
Indeed, the MA17 rate includes contributions from all possible low energy resonances ($E_x$=11.1-11.25 MeV), all of which are overwhelmed by the ($\alpha,n$) strength and the $(\alpha,\gamma)$ is negligible. 
In particular, the lowest resonance at $E_x$=11.11 MeV is the main source of the enhanced $^{22}$Ne($\alpha,n$)$^{25}$Mg rate. 

Compared to the rates given in TA16, our recommended rates produce up to a factor of 3 lower s-process abundances. However, the two sets of s-process calculations are consistent within the rate uncertainties given in  the two studies.
On the other hand, we obtain a reduction up to a factor of 10 compared to the results using the LO12 rates. In this case, the variation between the two sets of abundances is not compatible with the errors given.  
In general, we may conclude that within the uncertainties of our new $^{22}$Ne+$\alpha$ rates, we 
obtain a significant reduction in the s-process contribution 
 to the galactic chemical evolution of elements between iron and the s-process peak of Sr, Y and Zr. The significance of these effects 
will be studied in a forthcoming paper. 


\subsection{Sensitivity to $\alpha$ cluster strength}
\label{ss:AstroB}



In order to provide a complementary approach to assess the impact of the $^{22}$Ne+$\alpha$ rates provided in the previous section,  
we also calculated isotropic abundances when varying strengths of  key resonances at $E_x=11.11$, 11.17, and 11.32 MeV. 
Starting with the reaction rates presented in this work, we made calculations assuming four different conditions: 1) change $\Gamma_n/\Gamma_\gamma$ of the 11.32 MeV resonance from 1.14 to 3.78; 
2) insert a strong $\alpha$ cluster state at 11.17 MeV with significant $\gamma$-ray decay probability;
3) increase $\Gamma_\alpha$ of the 11.11 MeV resonance to 13.5 neV;
and 4) assume the 11.11 MeV resonance does not exist at all. 
Each of these changes is intended to probe the impact of existing discrepancies in the literature, or areas where substantial uncertainty still exists. 

Specifically, condition 1 probes the impact of the discrepancies between the present work (together with Refs.~\cite{Ota2019, Jayatissa2019}) and direct $(\alpha,n)$ measurements. The former established an $(\alpha,n)$ strength of $42(11)$ $\mu$eV, while the latter have a weighted average of $140(30)$ $\mu$eV for the same. In both cases, the $(\alpha,\gamma)$ strength is taken to be $37(4)$ $\mu$eV.

Condition 2 probes the impact of a strong $\alpha$ cluster state at 11.17 MeV, as claimed by TA16 but absent in the present work and others \cite{Giesen1993, Jayatissa2019, Ota2019}. In this condition, we inserted an ``observed'' (not upper limit) $\alpha\gamma$ resonance at 11.17 MeV with $\omega\gamma_{(\alpha\gamma)} = 660 \pm 90$ neV---the largest possible strength reported in TA16. Note that we modified the energy of this resonance slightly compared to TA16, placing it at 11.171 MeV instead of 11.167 MeV. We did this to account for the findings of MA17, which suggest a state with large $\gamma$-decay probability at $E_x=11.171$ MeV. For the $(\alpha,n)$ reaction, we increased the upper-limit strength of this resonance to $\omega\gamma < 60$ neV, as reported by Jaeger \emph{et al}.~\cite{Jaeger2001}.

%

Together, conditions 3 and 4 probe the impact of uncertainties on the $\alpha$ width, and hence $\omega\gamma_{(\alpha,n)}$ of the $2^+$ resonance at 11.11 MeV. Condition 3 probes the impact of a stronger 11.11 MeV resonance, up to the $\omega\gamma < 13.5$ neV ($\Gamma_\alpha<2.7$ neV) limit used to determine the upper-limit rates reported in MA17 \cite{MassimiEmail} (note that MA17's limits on the resonance strength were extracted from Ref.~\cite{Jaeger2001}).
To evaluate the maximum possible impact of this resonance, we treated it as an ``observed'' resonance in the Monte Carlo calculations, with a strength of $\omega\gamma = 13.5$ neV (as opposed to an ``unobserved'' resonance with the strength sampled from a Porter-Thomas distribution with an upper limit of 13.5 neV). As such, this calculation evaluates the impact of there being a strong 11.11 MeV resonance, with a strength just below the observation threshold of Ref.~\cite{Jaeger2001}.
%

In contrast to condition 3, condition 4 probes the impact of there being a negligibly small $\alpha$ width for the 11.11 MeV resonance, to the point where it can be removed from the rate calculations entirely.

\begin{figure*}
        \centering
          \includegraphics[width=17cm]{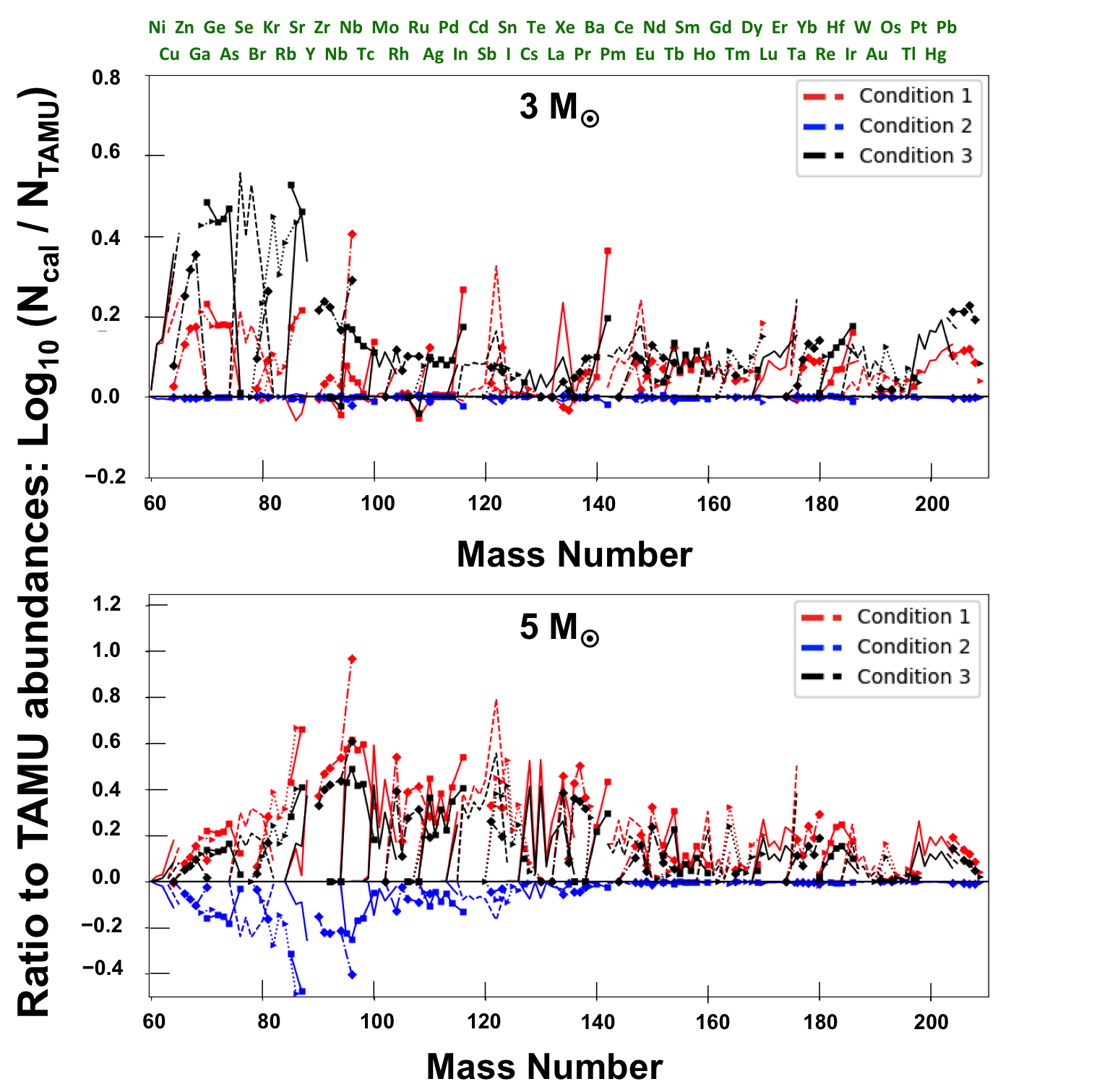}
        \caption{Ratios of calculated s-process abundances to TAMU abundances (see Figure~\ref{fig:Fig11}) using $^{22}$Ne($\alpha$,n) and $^{22}$Ne($\alpha$,$\gamma$) rates in which strength of some resonances are changed. See text for details.
        }\label{fig:Fig13}
\end{figure*}

\begin{figure*}
        \centering
          \includegraphics[width=17cm]{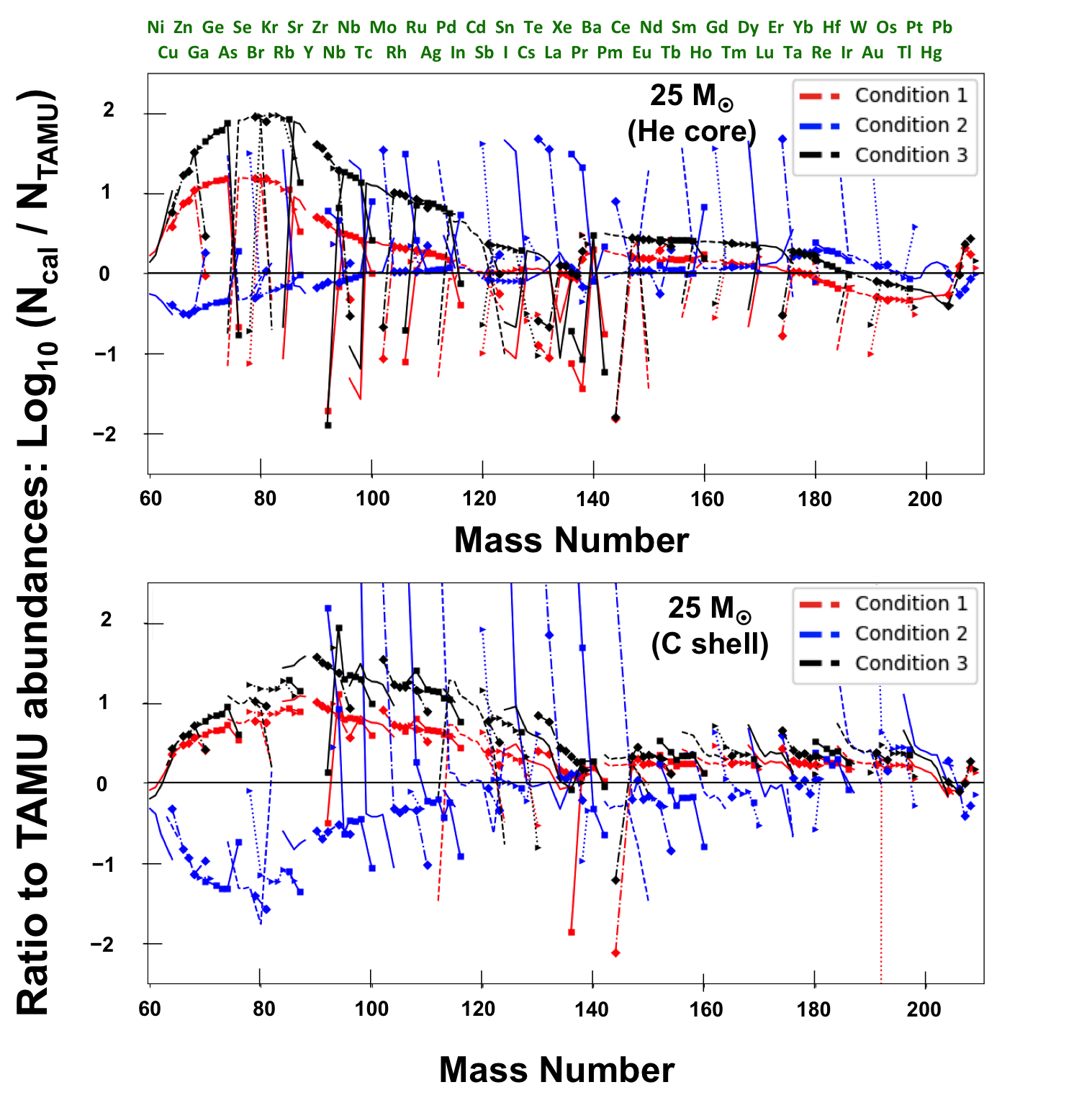}
        \caption{Ratios of calculated s-process abundances to TAMU abundances (see Figure~\ref{fig:Fig11b}) using $^{22}$Ne($\alpha$,n) and $^{22}$Ne($\alpha$,$\gamma$) rates in which strength of some resonances are changed. See text for details.
        }\label{fig:Fig13b}
\end{figure*}

\begin{figure*}
        \centering
          \includegraphics[width=13cm]{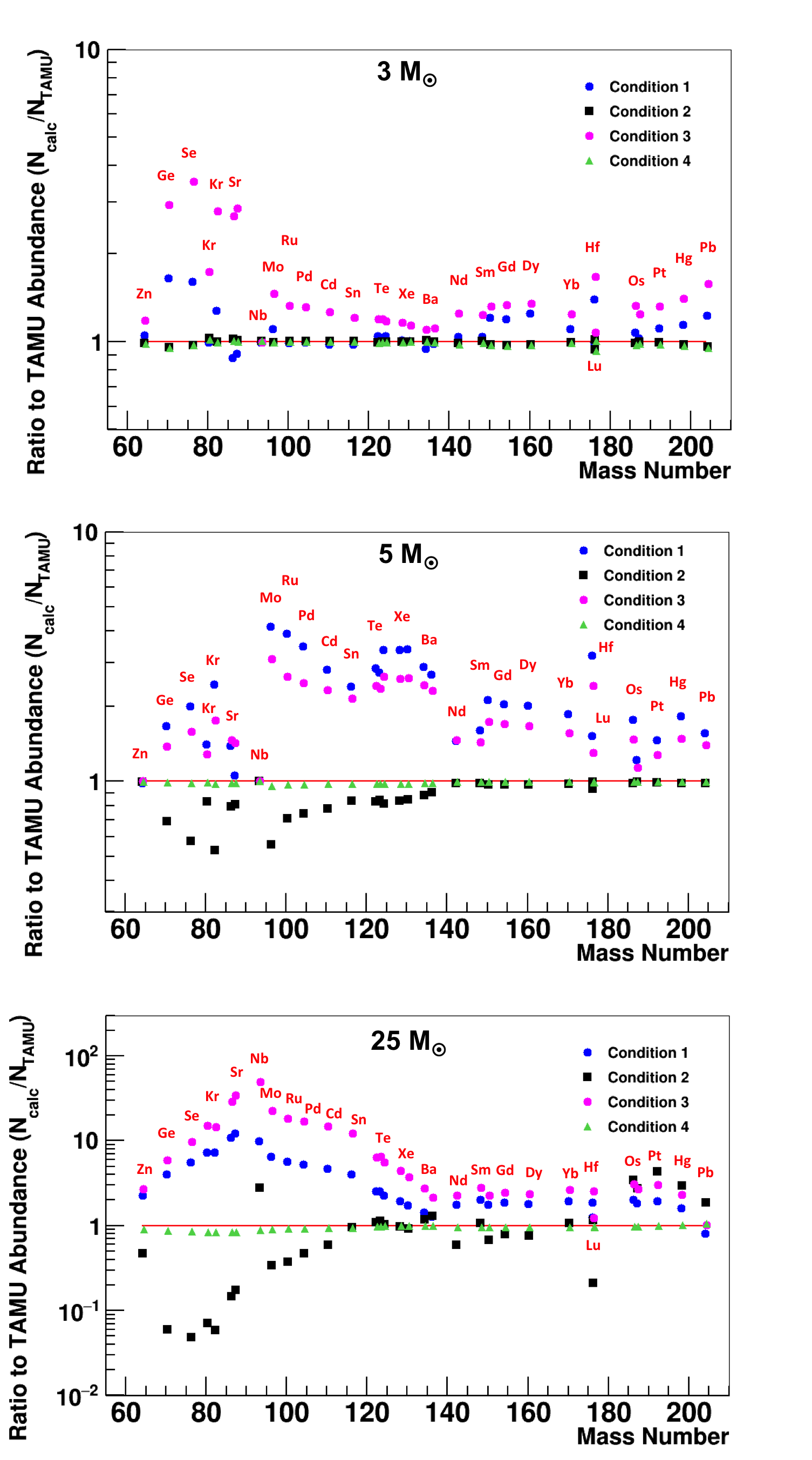}
        \caption{Calculated s-process abundance ratio to TAMU abundances for s-only nuclei (see Figure~\ref{fig:Fig13} and ~\ref{fig:Fig13b}) using $^{22}$Ne($\alpha$,n) and $^{22}$Ne($\alpha$,$\gamma$) rates in which strength of some resonances are changed. 
        Top, middle, and bottom panels correspond to 3, 5, and 25 (in the middle of the C shell burning) $M_\odot$ cases, respectively. 
        }\label{fig:Fig14}
\end{figure*}

Figure~\ref{fig:Fig13} and ~\ref{fig:Fig13b} show the resulting isotopic abundances when the reaction rates defined by conditions 1--3 are used. The abundances from each of the three conditions are plotted as a ratio to the baseline rate from OT20 (Texas A\&M (TAMU) abundances).
Figure~\ref{fig:Fig13}  shows abundances for 3 and  $5 M_\odot$ stars, and Figure~\ref{fig:Fig13b} shows $25 M_\odot$ abundances following both He-core and C-shell burning as indicated.
Figure~\ref{fig:Fig14} shows the same  abundance ratios as the previous two figures but for $s$-only nuclei exclusively. This figure also includes the calculations using the condition 4 rate. These are nearly indistinguishable from the OT20 results and thus were excluded from the previous two plots.


For the 3 $M_\odot$ model, the abundances are mainly sensitive to the changes in condition 3. The sensitivity to condition 1 is minor (25\% increase on average) and the impact of condition 2 is negligible.
The insensitivity to condition 2 mirrors the indistinguishable results between the LO12 and TA16 rates in the previous section (Figures~\ref{fig:Fig11} and \ref{fig:Fig12}). Evidently, the abundances for the 3 $M_\odot$ model are insensitive to a strong $(\alpha, \gamma)$ resonance at ${\sim}11.17$ MeV.
The minor sensitivity to the condition 1 rates  is limited to  $A<$90 nuclides, which again is consistent with the abundance differences from the previous section when comparing our rates and those of LO12 and TA16, which used the higher 11.32 MeV strength established in direct measurements. 
The large sensitivity to condition 3---increasing the 11.11 MeV resonance strength to MA17's upper limit---mirrors the previous section's differences between the present rates and those of MA17. 
The complete insensitivity to the changes of condition 4---removing the 11.11 MeV resonance---demonstrates that the strength of this resonance is already well constrained when considering its impact on $3 M_\odot$ stars.

For the 5 $M_\odot$ model, conditions 1 and 3 both result in a large increase in predicted abundances. Both rate changes have as much as a factor 3--4 influence on the abundances for certain elements, e.g.\ Mo. 
For this model, condition 2 decreases the predicted abundances by a factor of two at most. 
This reduction is expected because the increased $(\alpha,\gamma)$ rate, resulting from the strong 11.17 MeV resonance, competes with $(\alpha, n)$ for fuel. Hence, the stronger $(\alpha, \gamma)$ rate reduces neutron production via  ${}^{22}\mathrm{Ne}(\alpha,n)$.
The increase in the 11.11 MeV resonance strength (condition 3) again increases abundances substantially (although to a lesser degree than the $3 M_\odot$ model). The removal of the 11.11 MeV resonance entirely (condition 4) again results in negligible abundance changes.

For the 25 $M_\odot$ model, the largest sensitivity is observed for condition 3, similar to the other models, although now the impact is as large as a factor ${\sim} 100$ for elements near $A=100$. 
Condition 1 leads to up to a factor ${\sim} 10$ increase in production for the lighter nuclides. 
The impact of condition 2 is significant, leading to a large (factor $\lesssim 50$) decrease in abundances for elements near $A=80$. 
 Condition 4 again has a minor impact, although a perceptible decrease (factor ${\sim} 1.25$) in abundances is now present for the lightest nuclides.

\subsection{Summary of Astrophysical Implications and Suggested Future Nuclear Physics Studies}
\label{ss:AstroC}

The present study demonstrated that the upper limit on the strength of the 11.11 MeV resonance established in OT20 is sufficient to remove any observable impact of this state on $s$-process nucleosynthesis. However, as shown by our condition 3 calculations, if this state is substantially stronger than suggested in OT20, up to the 13.5 neV limit, it will have a dramatic influence on $s$-process nucleosynthesis. Such a scenario does not seem likely and would be inconsistent with the expected Porter-Thomas distribution of strengths. However, given the potential importance of this state to $s$-process nucleosynthesis, we encourage future studies that are specifically targeted at constraining its $\alpha$ width.

The second largest uncertainty in the $s$-process predictions appears to come from the 11.32 MeV resonance. 
The present ${\sim} 40 \mu$eV $(\alpha,n)$ resonance strength of this state has been established in two independent experiments \cite{Ota2019, Jayatissa2019}. However, this is a significant departure from the factor $3$--$4$ larger strength established in direct measurements---one that has a strong observable impact on $s$-process nucleosynthesis. As as result, we also encourage future experiments targeted at corroborating the results of Refs.~\cite{Ota2019,Jayatissa2019}.


Finally, we note that the impact of a strong $(\alpha,\gamma)$ resonance at 11.17 MeV appears to be relatively minor, limited to a reduction in the synthesis of $A \lesssim 100$ nuclides for 25 $M_\odot$ stars. Combined with the good agreement of  recent measurements that this state has a negligable $\alpha$ strength (e.g., Refs.~\cite{Ota2019, Jayatissa2019,Lotay2019}), we tentatively conclude that this state is not important for the $s$-process.

%

\section{Summary}

Natural parity states in $^{26}$Mg  were studied from $E_x$=4--12 MeV using the $^{22}$Ne($^6$Li, $d$) reaction in inverse kinematics. Coincidence tagging with the ${}^{25,26}$Mg recoils identified the first-order decay channels (neutron or $\gamma$ ray) of the populated states. The resulting spectra were interpreted with the help of coincident $\gamma$-ray measurements and the results of past $^{22}$Ne($^6$Li, $d$) measurements. This information was used to identify the most probable states contributing to the deuteron-reconstructed excitation-energy spectrum, which did not have sufficient resolution to separate all states populated in the reaction. Spins and relative spectroscopic factors were assigned to the identified states through comparison of angular distributions with DWBA calculations. The $^{22}$Ne($^6$Li,$t$) reaction was also analyzed. States likely to be strongly populated in this reaction were identified, and ($^6$Li,$t$) spectroscopic factors were reported for the first time.

The analysis of the ($^6$Li, $d$) data from this experiment, in the region of interest for the astrophysical $s$-process, led to the construction of new Mote-Carlo reaction rates for the stellar ${}^{22}\mathrm{Ne}(\alpha,n)$ and ${}^{22}\mathrm{Ne}(\alpha,\gamma)$ reactions \cite{Ota2019}. These rates were used to calculate predicted $s$-process abundances for a range of elements, using three different stellar models with $M/M_\odot = 3$, 5, and 25 and $\mathrm{Z}=0.02$. 
These calculations show the impact of the 
$^{22}$Ne+$\alpha$ reaction rates 
on the s-process abundances, using constraints from presently-available nuclear physics information. 
Using the new rates established by the results of the present experiment, 
 we observe a reduction in the overabundance of certain elements of up to a factor of 3 and 10 compared to the earlier rates published in Talwar \emph{et al}. (TA16) \cite{Talwar2016} and Longland \emph{et al}. (LO12) \cite{Longland2012}, respectively. 
These difference are mainly observed in mass range $A=60$--90, for all three stellar mass models. 

We also used our stellar models to investigate the impact on $s$-process nucleosynthesis of outstanding uncertainties or literature discrepancies on selected, key ${}^{22}\mathrm{Ne} + \alpha$ resonances.
These calculations used the present Monte-Carlo rates as a baseline and subsequently varied strengths of the resonances at $E_x = 11.11$, 11.17, and 11.32 MeV. The results highlight the strong astrophysical impact of the factor ${\sim} 3$ reduction in the $E_x = 11.32$ MeV $(\alpha,n)$ strength, which resulted from the present experiment.
The calculations also established that the discrepancies between the present work (along with others \cite{Jayatissa2019, Giesen1993, Longland2012}) and TA16 concerning the presence of a strong $(\alpha,\gamma)$ resonance at 11.17 MeV are relatively minor except for the 25 $M_\odot$ stars. 
Finally, the calculations established that the presently-determined upper limit on the strength of the $2^+$ resonance at 11.11 MeV is sufficient to constrain $s$-process nucleosynthesis across all models. At the same time, all three models predict a strong enhancement in s-process abundances if a resonance is found just below the observation threshold of Ref.~\cite{Jaeger2001}.
 This highlights the significant importance of this resonance to s-process nucleosyntiesis. All together, the calculations point to the 11.11 and 11.32 MeV resonances as being the most important for the s-process. As a result, we encourage future experiments targeted at independently corroborating the strengths of these resonances established by the present experiment.


\begin{acknowledgments}
We express our thanks to the technical staff at the Texas A\&M University Cyclotron Institute. Financial support for this work was provided by the US Department of Energy, award Nos.\ DE-FG02-93ER40773 and DE-SC0018980, the US National Nuclear Security Administration, award No.\ DE-NA0003841.  JAT, WNC., and GL. acknowledge support from the UK STFC, award no.\ ST/L005743/1.  SO acknowledges support from the TAMU CIRD fund.
MP acknowledges support from NuGrid, JINA-CEE (NSF Grant PHY-1430152),  STFC (through the University of Hull’s Consolidated Grant ST/R000840/1), and ongoing access to {\tt viper}, the University of Hull High Performance Computing Facility. MP also acknowledges the support from the "Lend{\"u}let-2014" Programme of the Hungarian Academy of Sciences (Hungary). We also acknowledge the ChETEC COST Action (CA16117), supported by the European Cooperation in Science and Technology, the CheTEC-INFRA (European Union’s Horizon 2020 research and innovation program, grant agreement No 101008324) and IReNA (USA National Science Foundation under Grant No. OISE-1927130). 
\end{acknowledgments}



\bibliographystyle{unsrt}
\bibliography{Mg26And25_2019}

\end{document}